\documentclass[useAMS,usenatbib]{mn2e}
\usepackage{graphicx}
\usepackage{amssymb}
\usepackage{bm}

\usepackage{color} 
\def\jref@jnl#1{{\rm#1}}

\def\aj{\jref@jnl{AJ}}                   
\def\araa{\jref@jnl{ARA\&A}}             
\def\apj{\jref@jnl{ApJ}}                 
\def\apjl{\jref@jnl{ApJ}}                
\def\apjs{\jref@jnl{ApJS}}               
\def\ao{\jref@jnl{Appl.~Opt.}}           
\def\apss{\jref@jnl{Ap\&SS}}             
\def\aap{\jref@jnl{A\&A}}                
\def\aapr{\jref@jnl{A\&A~Rev.}}          
\def\aaps{\jref@jnl{A\&AS}}              
\def\azh{\jref@jnl{AZh}}                 
\def\baas{\jref@jnl{BAAS}}               
\def\jrasc{\jref@jnl{JRASC}}             
\def\memras{\jref@jnl{MmRAS}}            
\def\mnras{\jref@jnl{MNRAS}}             
\def\pra{\jref@jnl{Phys.~Rev.~A}}        
\def\prb{\jref@jnl{Phys.~Rev.~B}}        
\def\prc{\jref@jnl{Phys.~Rev.~C}}        
\def\prd{\jref@jnl{Phys.~Rev.~D}}        
\def\pre{\jref@jnl{Phys.~Rev.~E}}        
\def\prl{\jref@jnl{Phys.~Rev.~Lett.}}    
\def\pasp{\jref@jnl{PASP}}               
\def\pasj{\jref@jnl{PASJ}}               
\def\qjras{\jref@jnl{QJRAS}}             
\def\skytel{\jref@jnl{S\&T}}             
\def\solphys{\jref@jnl{Sol.~Phys.}}      
\def\sovast{\jref@jnl{Soviet~Ast.}}      
\def\ssr{\jref@jnl{Space~Sci.~Rev.}}     
\def\zap{\jref@jnl{ZAp}}                 
\def\nat{\jref@jnl{Nature}}              
\def\iaucirc{\jref@jnl{IAU~Circ.}}       
\def\aplett{\jref@jnl{Astrophys.~Lett.}} 
\def\apspr{\jref@jnl{Astrophys.~Space~Phys.~Res.}}
\def\bain{\jref@jnl{Bull.~Astron.~Inst.~Netherlands}} 
\def\fcp{\jref@jnl{Fund.~Cosmic~Phys.}}  
\def\gca{\jref@jnl{Geochim.~Cosmochim.~Acta}}   
\def\grl{\jref@jnl{Geophys.~Res.~Lett.}} 
\def\jcp{\jref@jnl{J.~Chem.~Phys.}}      
\def\jgr{\jref@jnl{J.~Geophys.~Res.}}    
\def\jqsrt{\jref@jnl{J.~Quant.~Spec.~Radiat.~Transf.}}
\def\memsai{\jref@jnl{Mem.~Soc.~Astron.~Italiana}}
\def\nphysa{\jref@jnl{Nucl.~Phys.~A}}   
\def\physrep{\jref@jnl{Phys.~Rep.}}   
\def\physscr{\jref@jnl{Phys.~Scr}}   
\def\planss{\jref@jnl{Planet.~Space~Sci.}}   
\def\procspie{\jref@jnl{Proc.~SPIE}}   

\title[Orbital dynamics of a double neutron star system around a black hole]{Complex orbital dynamics of a double neutron star system revolving around a massive black hole}
\author[Grant N. Remmen and Kinwah Wu]{Grant N. Remmen$^{1,2,3}$\thanks{E-mail: gremmen@caltech.edu (GNR); kw@mssl.ucl.ac.uk (KW)} 
and Kinwah Wu$^{1}$\footnotemark[1]
\\
$^{1}$Mullard Space Science Laboratory, University College London, Holmbury St.\,Mary, Dorking, Surrey RH5 6NT    \\
$^{2}$Minnesota Institute for Astrophysics, School of Physics and Astronomy, University of Minnesota, 
Minneapolis, MN 55455, USA \\ 
$^{3}$Division of Physics, Mathematics and Astronomy, California Institute of Technology, Pasadena, CA 91125, USA} 

\begin{document}
\onecolumn

\date{Accepted 2013 *** ***. Received 2012 *** ***; in original form 2012 *** ***}

\pagerange{\pageref{firstpage}--\pageref{lastpage}} \pubyear{2012}

\maketitle

\label{firstpage}

\begin{abstract} 

We investigate the orbital dynamics of hierarchical three-body systems 
  containing a double neutron star system orbiting around a massive black hole. 
These systems show complex dynamical behaviour 
  because of relativistic coupling between orbits of the neutron stars in the double neutron star system 
  and the orbit of the double neutron star system around the black hole.     
The orbital motion of the neutron stars around each other drives a loop mass current, 
  which gives rise to gravito-magnetism. 
Generally, gravito-magnetism involves a rotating black hole. 
The hierarchical three-body system that we consider is an unusual situation in which black hole rotation is not required. 
Using a gravito-electromagnetic formulation, we calculate the orbital precession and nutation of the double neutron star system. 
These precession and nutation effects are observable, 
  thus providing probes to the spacetime around black holes 
  as well as tests of gravito-electromagnetism in the framework of general relativity.  
  
\end{abstract}


\section{Introduction}  

There is strong evidence that the Galactic Centre contains a black hole 
  with mass $M \approx 4 \times 10^{6}{\rm M}_{\odot}$ \citep{Gillessen09}.  
X-ray observations \citep[e.g.][]{Muno09} indicate that 
  the Galactic central region 
  contains a large number of stellar remnants.  
There have been studies suggesting that      
  about 20,000$-$40,000 stellar-mass black holes \citep{Miralda00} 
  and thousands of neutron stars \citep{Freitag06,Wharton12} are 
  residing in the region. 
The presence of such a large black hole population requires that the Galactic Centre is dynamically relaxed.  
However, recent studies \citep{Buchholz09, Do09, Bartko10} have indicated otherwise, 
  implying that there may well be much fewer stellar-mass black holes in the Galactic Centre \citep{Merritt10, Antonini12}.     
The estimate of $\sim 10^{3}$ neutron stars    
  is based on the models assuming the presence of a cluster of $\sim 10^{4}$ stellar-mass black holes in the same region. 
Without this stellar-mass black hole cluster, the number of neutron stars would be higher \citep{Freitag06}.      
A half dozen pulsars have already been discovered in the central parsec of Sgr A$^{*}$ \citep{Deneva09, Macquart10}, 
  showing the evidence of the neutron star population in the Galactic Centre.     
Centres of galaxies with spheroids similar to that of our Galaxy are expected to host thousands or more neutron stars.     
Large elliptical galaxies have massive stellar spheroids and their neutron star populations are scaled accordingly.  
Some of these neutron stars would eventually fall into the massive black holes, 
  forming EMRI (extreme mass-ratio inspiral) systems, 
  which are of great interest in gravitational wave astrophysics \citep[see][]{Sathyaprakash09}.  
 
Many neutron stars reside in binary systems, e.g.\ X-ray binaries and pulsar binaries \citep[see][]{Liu06, Liu07, Lorimer08}.   
Some neutron stars also form pairs, known as double neutron star (DNS) systems, e.g. PSR~B1913+16.  
DNS systems are expected to be rare.      
However, more than a dozen DNS systems have already been found \citep[see][]{Lorimer08}. 
The discovery of the DNS system B2127+11C, 
  which is probably being ejected from the core of the globular cluster M15 \citep{Prince91},  
  suggests that dense stellar environments could produce DNS systems efficiently. 
The progenitors of DNS systems are massive star binaries.     
A recent study \citep{Sana12} showed an unexpectedly large fraction of massive stars are in binary systems  
  with over 70\% of all massive stars transferring material to their companions.  
Finding such a large fraction of massive star binaries with substantial mass exchange   
  clearly indicates that the evolution of massive binary stars is far from trivial  
  and that DNS systems may be more abundant than previously thought.  

Given a substantial number of neutron stars in the central regions of galaxies, 
  some of them would form DNS systems.   
When a DNS system sinks deeply toward the massive central black hole of its host galaxy,  
    it would end up in a close orbit around the black hole before the final in-spiral.  
DNS systems are 
detectable if they contain a pulsar. 
However, detecting pulsars and binary pulsars in the centres of galaxies, e.g.\ the Galactic Centre,  
  is a great technical challenge in observation \citep[see][]{Bates11}.   

The dynamics of strongly bound gravitational systems often harbor interesting complex phenomena.  
Studies \citep[e.g.][]{Singh05, Singh08} have shown that a fast spinning neutron star orbiting around a black hole 
  will exhibit complex spin precession due to various relativistic couplings.   
For pulsars, the spin precession of neutron stars orbiting around black holes 
  will be manifested in the variation of their pulse emission \citep{Wex99, Liu12}.    
Pulsars are therefore useful experimental probes of relativistic spin-orbit interactions.    
Hierarchical three-body systems, such as compact binaries orbiting around massive black holes, 
  are known to show complex dynamical behaviours \citep[see e.g.][]{Antonini12}. 
Tightly bound DNS systems are a special subclass of such hierarchical three-body systems. 
Like pulsars, DNS systems are also spinning objects, 
  and hence they would experience similar spin-orbit interactions as they revolve around a black hole.  
However, DNS systems are not point gyroscopes, unlike pulsars, 
  and the subjection of a DNS system to tidal interactions in the gravitational field of the black hole gives rise to richer dynamical behaviours.   
   
In this study, we explore the complex orbital dynamics in the hierarchical three-body system 
  consisting of a DNS system orbiting around a massive black hole.  
We consider the tightly bound DNS systems, 
  which have orbital periods much shorter than the periods with which they revolve around the black holes.  
The focus of this work is on the complex dynamics 
  arising from the interaction between the orbit of the two neutron stars 
  and the orbit of the DNS system around the massive black hole.  
We will leave the more complicated interaction between the orbit of the two neutron stars and the spin of the black hole to a future study. 
We organize the paper as follows. 
We first formulate the dynamical interaction between the internal and external orbits of the DNS system revolving around a black hole 
   and derive the corresponding effective potentials (\S 2).   
We recast the formulation in terms of Euler angles for the orientation of the (internal) orbital plane of the neutron stars in the DNS system  
   and construct the Lagrangian, complete with terms corresponding to Coriolis and centrifugal forces for the orbiting reference frame. 
Then we derive the equations of motion that govern the orbital dynamics of the system (\S 3). 
We use two special cases to illustrate the precession and nutation behaviours, 
  and finally we present a full numerical solution (\S4).  
The physical and astrophysical implications are discussed.

\section{Interactions and Effective Potentials} 

\begin{figure*}
\noindent 
\begin{centering}
\vspace*{0.2cm}
\includegraphics[width=9.5cm]{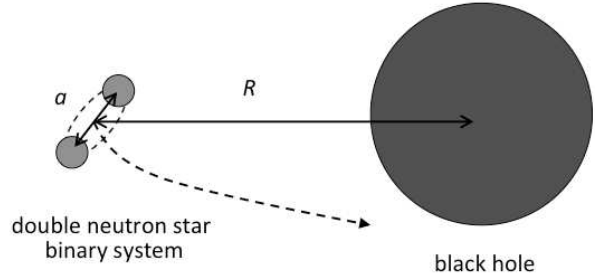} 
\vspace*{-0.25cm}
\end{centering}
\caption{A schematic illustration of the configuration of a tightly bound double neutron star (DNS) system  
  orbiting around a massive black hole (not to scale). 
  The orbital separation of the two neutron stars is $a$ and the radius of the orbit of the DNS system around the black hole is $R$. 
  The orbit of the neutron stars in the DNS system and the orbit of the DNS system around the black hole 
     are not necessarily co-planar. 
    } 
\label{DNS} 
\end{figure*}

The configuration of the hierarchical DNS and black hole system is shown in Fig.~\ref{DNS}.    
We treat the black hole and the neutron stars as point masses.  
As neutron stars have very similar masses, around 1.5~M$_{\odot}$ \citep{Lattimer11}, 
  we consider that the two neutron stars have equal masses $m$.   
The mean orbital separation between the neutron stars is $a$,  
  and the distance of the neutron star to their centre of mass is therefore $a/2$. 
The DNS system is orbiting about a massive black hole with a mass $M$ substantially larger than the that of the neutron star mass, i.e. $M\gg m$.   
The orbital radius of the DNS system is $R$. 
Moreover, $R\gg a$. 
As the primary aim of this study is to demonstrate that complex behaviours can arise from orbital coupling in the DNS systems 
  even in very simple configurations, 
  we avoid unnecessary complications. 
We consider Schwarzschild spacetime for the black hole and zero orbital eccentricity for the DNS system.  
  
The orbital angular frequency of the neutron stars in the DNS system is given by $\Omega_{\rm ns}  = \sqrt{Gm/2a^{3}}$, 
  and the orbital angular frequency of DNS system around the black hole is given by $\Omega_{\rm bh}=\sqrt{GM/R^{3}}$.     
The condition of $R/a \gg  (2M/m)^{1/3}$ implies that $\Omega_{\rm ns} \gg \Omega_{\rm bh}$.  
This condition is easily satisfied with the systems of our interest, 
  so the DNS system can be considered as tightly bound.  
The fast orbital revolution of the neutron stars around each other effectively generates a looped mass current.    
On the time scales $t \gg 2\pi / \Omega_{\rm bh}$,     
  the continuous approximation is applicable   
  and the looped mass current effectively makes the tightly bound DNS system into a relatively rigid rotating mass ring.    
It is important to note that a two-body DNS system is not identical a rigid rotating mass ring, 
  as the orbit of the former is maintained by gravity and the shape of the ring is determined by the constraint force.  
However,  in the Lagrangian formulation for the dynamical analysis of a rigid ring in an external field, 
  consideration of constraint forces within the ring need not be included explicitly.  
In essence, the distance $a/2$ of each differential mass element from the centre of the ring 
  forms a holonomic constraint \citep[see][]{Goldstein50, Fowles05}, 
  with the result that any displacement from this rigid configuration is an ignorable coordinate in the Lagrangian \citep{Fowles05}.  
Thus, the justification of the the rotating mass ring representation for the DNS systems 
  gives us tremendous simplification 
  in the mathematics when carrying out the analysis of the dynamical coupling 
  between the internal orbit of the neutron stars in the DNS system and its external orbit of the DNS system around the massive black hole.


\subsection{Tidal interaction}  

We start the derivation of the effective potential by considering the tidal force on the DNS system. 
The tidal force between freely falling objects moving along two
  adjacent geodesics with separation \mbox{\boldmath${\eta}$} may be derived from the Riemann curvature tensor 
  in the local, orthonormal, inertial frame \citep{Misner77, Rindler06}: 
\begin{equation}
\frac{D^{2}\eta^{j}}{d\tau^{2}}=-R_{\tau j\tau k}\eta^{k}\,,\label{tidaltensor}
\end{equation}
  where $d\tau$ denotes the unit of proper time at that spacetime location and $D$ is the covariant derivative operator. 
For the Schwarzschild metric,  
\begin{eqnarray}
\frac{D^{2}\eta^{r}}{d\tau^{2}} & = & \frac{2GM}{r^{3}}\eta^{r}\   ; 
  \label{tidalr}   \\
\frac{D^{2}\eta^{\theta}}{d\tau^{2}} & = & -\frac{GM}{r^{3}}\eta^{\theta}\  ; 
\label{tidaltheta}   \\
\frac{D^{2}\eta^{\phi}}{d\tau^{2}} & = & -\frac{GM}{r^{3}}\eta^{\phi}\     
\label{tidalphi}
\end{eqnarray}  
\citep[see also][]{Fang83}. 
The tidal force stretches objects in the radial direction and compresses them within the plane normal to the radial.  
As derived in \citet{Ohanian94} for \mbox{\boldmath${\eta}$} suitably small 
  as in the case of the DNS and black hole system considered here, where $a\ll R$, 
  Eq.~\ref{tidaltensor} may be expressed in the Schwarzschild time coordinate $t$. 
Hence, Eqs.~\ref{tidalr} through~\ref{tidalphi}
  may be recast in terms of an effective potential, seen by a distant observer, acting relative to the DNS system's centre of mass:   
\begin{eqnarray}
   \Phi & = & \frac{1}{2}\frac{GM}{R^{3}}\left[\left(\eta^{\theta}\right)^{2}+\left(\eta^{\phi}\right)^{2}-2\left(\eta^{r}\right)^{2}\right]  \nonumber \\ 
          & =& \frac{1}{2}\frac{GM}{R^{3}}\left[\left(\frac{a}{2}\right)^{2}-3\left(\eta^{r}\right)^{2}\right],\label{tidalPhi}
\end{eqnarray}
    since $\left(\eta^{r}\right)^{2}+\left(\eta^{\theta}\right)^{2}+\left(\eta^{\phi}\right)^{2}=\left(a/2\right)^{2}$.

\begin{figure*} 
\vspace*{0.1cm}
\noindent  
\begin{centering}
\includegraphics[width=5cm]{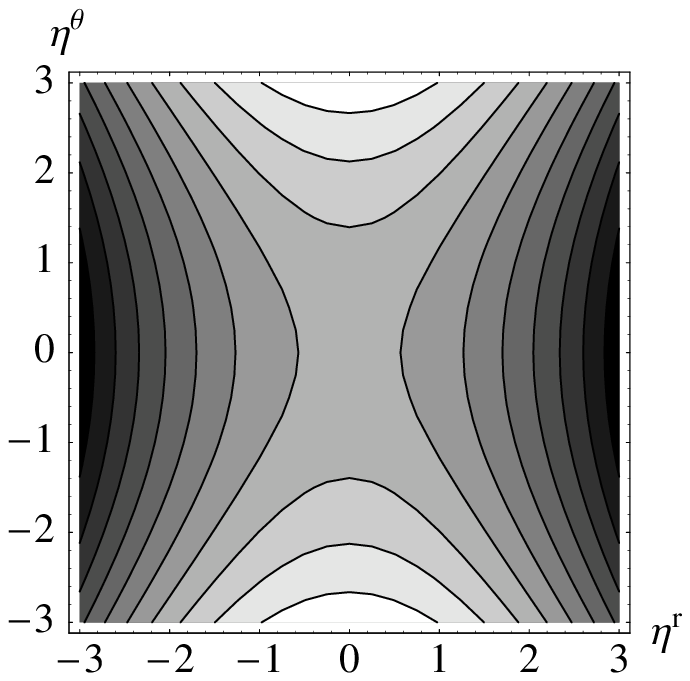} \hspace*{0.4cm}
\includegraphics[width=5cm]{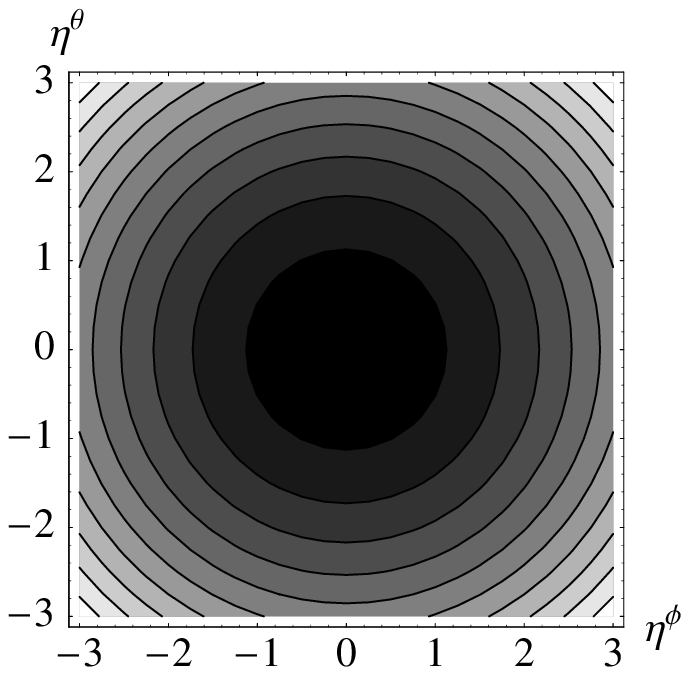} \hspace*{0.2cm}
\includegraphics[width=4.6cm]{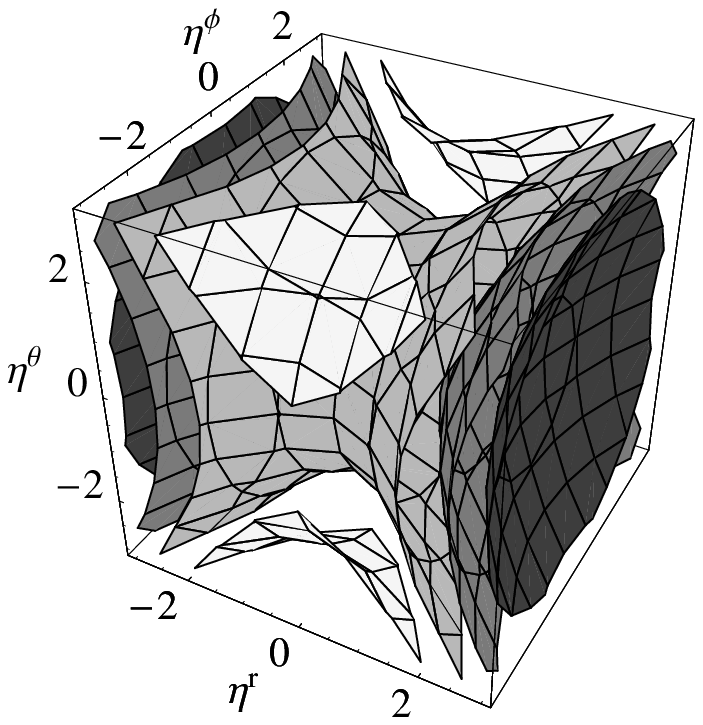}  
\vspace*{0.3cm}
\end{centering}
\caption{Potential contours, illustrating slices through the effective potential $\Phi$ of the Schwarzschild tidal force. 
  Potential increases from dark to light shading. 
  Left: Hyperbolic contours, of a saddle-like surface, in a plane containing the radial vector. 
  Centre: Circular contours, of a paraboloidal surface, in the plane normal to the radial vector. 
  Right: Hyperboloidal equipotential surfaces in three dimensions.
} 
\label{tidalfig} 
\end{figure*}

The equipotential surfaces of $\Phi$ are hyperboloids, 
   with a saddle point at the centre of mass of the DNS system (the mass ring),  
  as illustrated in Fig.~\ref{tidalfig}.  
This potential is derived in the weak-field, slow-motion approximation. 
It is a special case of the gravitational quadrupole potential as that shown in \cite{Barker75a}. 
Here we have shown that for a DNS system orbiting a black hole 
  the potential in the mass ring representation may be simply derived 
  as following immediately from the equations for the tidal force.  

The potential energy $V_{\mathrm{tidal}}$ of the mass ring as a result of its orientation within the tidal field 
  is given by integrating $\Phi$ over each mass element in the mass ring, i.e.\ 
\begin{equation}
   V_{\mathrm{tidal}}=\oint_{\mathrm{ring}}dl\,\lambda\Phi \    , 
\label{tidalen}
\end{equation}
   where $\lambda=2m/\pi a$. 
Let the plane of the ring be defined by normal unit vector \mbox{\boldmath${\hat{\mathrm{n}}}$}, 
  with zenith and azimuth coordinates $\theta$ and $\phi$. 
Parameterizing the mass elements of the ring in terms of an angular element $d\chi=2\, dl/a$ permits 
  the expression of \mbox{\boldmath${\eta}'$} for a particular element
  via coordinate rotation of the vector $\left(\cos\chi,\;\sin\chi,\;0\right)^{T}$,
  which describes a ring in the $x$-$y$ plane, into the plane of the ring:
\begin{eqnarray}
    {\bm{\eta}}' & = & \left(\begin{array}{c}
        \eta^{r}  \\
       \eta^{\phi}  \\
      -\eta^{\theta}
    \end{array}\right) \nonumber \\ 
  & =& \left(\frac{a}{2}\right)  \left(\begin{array}{ccc}
     \cos\theta\cos\phi' &  -\sin\phi' & \sin\theta\cos\phi'   \\
        \cos\theta\sin\phi' & \cos\phi' & \sin\theta\sin\phi'   \\
       -\sin\theta & 0 & \cos\theta
    \end{array}\right)\left(\begin{array}{c}
    \cos\chi\\
   \sin\chi\\
    0
    \end{array}\right)\nonumber \\
 & = & \left(\frac{a}{2}\right)\left(\begin{array}{c}
   \cos\theta\cos\phi'\cos\chi-\sin\phi'\sin\chi\\
    \cos\theta\sin\phi'\cos\chi+\cos\phi'\sin\chi\\
   -\sin\theta\cos\chi
    \end{array}\right) \ ,
\end{eqnarray}
which gives 
\begin{eqnarray}
\left(\eta^{r}\right)^{2} & = & \left(\frac{a}{2}\right)^{2}\Bigl(\cos^{2}\theta\cos^{2}\phi'\cos^{2}\chi+\sin^{2}\phi'\sin^{2}\chi  
   - 2\cos\theta\cos\phi'\sin\phi'\cos\chi\sin\chi\Bigr) \  . 
\label{etalength}
\end{eqnarray}
The potential at a particular element is thus 
\begin{eqnarray}
   \Phi & = & \frac{1}{2}\frac{GM}{R^{3}}\left(\frac{a}{2}\right)^{2}\Bigl[1-3\Bigl(\cos^{2}\theta\cos^{2}\phi'\cos^{2}\chi+\sin^{2}\phi'\sin^{2}\chi  
   - 2\cos\theta\cos\phi'\sin\phi'\cos\chi\sin\chi\Bigr)\Bigr] \ .
\label{tidalintegrand}
\end{eqnarray}
Here the azimuth angle is expressed in co-rotating coordinates, $\phi'=\phi-\Omega t$,
   where $\Omega\ (=\Omega_{\rm bh} = \sqrt{GM/R^{3}})$ is the angular speed of the orbit in coordinate time. 
Hence, $\phi'$ is the angle between the radial  vector from the central gravitating body (the massive black hole) to the ring (the DNS system), 
   \mbox{\boldmath${\mathrm{R}}$},
   and the projection of \mbox{\boldmath${\hat{\mathrm{n}}}$} 
   into the plane defined by  \mbox{\boldmath${\mathrm{R}}$} and 
   \mbox{\boldmath${\mathrm{R}}$}$\times$\mbox{\boldmath${\mathrm{L}}$},
   where  \mbox{\boldmath${\mathrm{L}}$} is the orbital angular momentum.
That is, when \mbox{\boldmath${\mathrm{L}}$} is oriented along the \mbox{\boldmath${\hat{\mathrm{z}}}$} direction, 
   so that the orbit about the massive body is in the $x$-$y$ plane, 
    $\phi'$ is the angle between \mbox{\boldmath${\mathrm{R}}$} and
   the projection of $\bm{\hat{\mathrm{n}}}$ into the $x$-$y$ plane. 
Integrating yields the total orientational potential energy of the ring:
\begin{eqnarray}
V_{\mathrm{tidal}} & = & \frac{1}{2\pi}\frac{GMm}{R^{3}}\left(\frac{a}{2}\right)^{2}\int_{0}^{2\pi}d\chi\Bigl[1-3\bigl(\cos^{2}\theta\cos^{2}\phi'\cos^{2}\chi 
   +\sin^{2}\phi'\sin^{2}\chi -2\cos\theta\cos\phi'\sin\phi'\cos\chi\sin\chi\bigr)\Bigr]\nonumber \\
    & = & \frac{GMm}{R}\left(\frac{a}{2R}\right)^{2}\left\{ 1-\frac{3}{2}\left[\cos^{2}\theta\cos^{2}\left(\phi-\Omega t\right) 
     +\sin^{2}\left(\phi-\Omega t\right)\right]\right\}\  . 
\label{tidalpotentialint}
\end{eqnarray} 


Note that Eq.~\ref{tidalpotentialint}, which we have derived completely classically, 
  may be compared with the orientational potential energy 
  from the Hamiltonian for the gravitational two-body problem in~\citet{Barker75b} 
  discussed in a semi-classical context in \citet{Barker75a} and \citet{Chan77}, 
  which gives the potential energy for a mass with quadrupole moment $J_{2}$. 
In terms of our notation, their expression is
\begin{equation}
   V_{\mathrm{tidal}}=\frac{GJ_{2}M\left(2m\right)}{2R^{3}}\left[\frac{3\left(\bm{\hat{\mathrm{n}}}\cdot\bm{\mathrm{R}}\right)^{2}}{R^{2}}-1\right] \ .
\label{tidalpotentialgeneral}
\end{equation}
For the thin ring, $J_{2}=\left(1/2\right)\left(a/2\right)^{2}$,
   so with 
$\bm{\mathrm{R}}=R\left(\bm{\hat{\mathrm{x}}}\cos\Omega t+\bm{\hat{\mathrm{y}}}\sin\Omega t\right)$
   and $\bm{\hat{\mathrm{n}}}=\bm{\hat{\mathrm{x}}}\sin\theta\cos\phi+\bm{\hat{\mathrm{y}}}\sin\theta\sin\phi+\bm{\hat{\mathrm{z}}}\cos\theta$,
$\bm{\hat{\mathrm{n}}}\cdot \bm{\mathrm{R}}/R
 =\sin\theta\cos\phi\cos\Omega t+\sin\theta\sin\phi\sin\Omega t=\sin\theta\left[\cos\phi\cos\left(-\Omega t\right)-\sin\phi\sin\left(-\Omega t\right)\right]=\sin\theta\cos\left(\phi-\Omega t\right)$.
Hence, Eq.~\ref{tidalpotentialgeneral} becomes 
\begin{eqnarray}
V_{\mathrm{tidal}} & = & \frac{1}{2}\frac{GMm}{R}\left(\frac{a}{2R}\right)^{2}\left[3\sin^{2}\theta\cos^{2}\left(\phi-\Omega t\right)-1\right] 
   \nonumber \\
 & = & \frac{GMm}{R}\left(\frac{a}{2R}\right)^{2}\left\{ 1-\frac{3}{2}\left[\cos^{2}\theta\cos^{2}\left(\phi-\Omega t\right)+\sin^{2}\left(\phi-\Omega t\right)\right]\right\}\  ,
\end{eqnarray}
identically the result obtained through the effective potential formalism
in Eq.~\ref{tidalpotentialint}.

\subsection{Orbital coupling}

We next derive the potential due to orbital interaction. 
The transport of vectors, such as angular momentum, along paths in curved spacetime gives rise to precessional effects. 
Within the Schwarzschild metric, 
  the transport of angular momentum vectors along circular orbits is subject to deSitter precession \citep{deSitter16}.   
That is, the orientation of the vector will precess with angular velocity 
  proportional to \mbox{\boldmath${\mathrm{r}}$}$\times$$\bm{\mathrm{v}}$,
  where \mbox{\boldmath${\mathrm{r}}$} and \mbox{\boldmath${\mathrm{v}}$} are
  the position and velocity of the test body carrying the angular momentum vector 
  \citep{Misner77, Rindler06}. 
This implies that when the DNS system revolves around the black hole, 
  the orbital angular momentum vector of the neutron stars about each other will gradually change its orientation.  
 
The deSitter precession can be derived geodesically,
   i.e.\ directly finding the deflection resulting from parallel transport,
   or equivalently, by converting the Schwarzschild metric to a rotating coordinate system. 
However, if a Lagrangian analysis within the Schwarzschild metric is required, 
  it is more calculationally useful and conducive to physical insight 
  to express the coupling between the angular momentum of the spinning body about its centre of mass ($\bm{\mathrm{S}}$), 
  and the orbital angular momentum ($\bm{\mathrm{L}}$)  as a potential energy, $V_{SL}$. 
Obviously, here $\bm{\mathrm{S}}$ corresponds to the summed orbital angular momentum of the neutron stars 
  in the centre of mass frame of the DNS,   
  and $\bm{\mathrm{L}}$ corresponds to the angular momentum of the DNS system's orbit around the black hole.  

  
We now show that the general relativistic orbital coupling between the DNS system and the black hole 
  can be derived using the gravito-electromagnetic (GEM) formalism.
Note that the results are also applicable for spin-orbit coupling in binary systems of arbitrary mass ratios.  
The GEM equations constitute a next-to-leading order approximation to the Einstein field equations \citep[see e.g.][]{Misner77,Punsly01}.  
They closely resemble the form of Maxwell's equations. 
In essence, the derivation of the gravitational spin-orbit interaction energy 
   is analogous to the derivation of the spin-orbit energy of the electron in the hydrogen atom \citep[see][]{Jackson99}. 

The GEM equations read
\begin{eqnarray} 
{\nabla}\cdot\bm{\mathrm{g}} & = & 4\pi G\rho_{\mathrm{g}}  \ ;  \label{GEMcoulomb} \\
{\nabla}\cdot\frac{\bm{\mathrm{h}}}{2} & = & 0 \ ;   \label{GEMdivh} \\
{\nabla}\times\bm{\mathrm{g}} & = & 
   -\frac{1}{c}\frac{\partial}{\partial t}\left(\frac{1}{2}\bm{\mathrm{h}}\right)\ ;  \label{GEMfaraday} \\
  {\nabla}\times\frac{\bm{\mathrm{h}}}{2} & = &
   \frac{4\pi G}{c}\bm{\mathrm{j}}_{\mathrm{g}}+\frac{1}{c}\frac{\partial\bm{\mathrm{g}}}{\partial t} \ , 
  \label{GEMamperemaxwell}
\end{eqnarray}
  where $\bm{\mathrm{g}}$ is the negative of the usual gravitational field strength, 
  $\bm{\mathrm{h}}$ is the gravito-magnetic field, 
  $\rho_{\mathrm{g}}$ is the mass density, 
  and $\bm{\mathrm{j}}_{\mathrm{g}}$ is the mass current density \citep{Punsly01,Ruggiero02,Mashhoon07}. 
The GEM Lorentz force law for an object of mass $m$ is 
\begin{equation}
\bm{\mathrm{F}}=-m\left(\bm{\mathrm{g}}+2\frac{\bm{\mathrm{v}}}{c}\times\bm{\mathrm{h}}\right)\ . 
 \label{GEMlorentz}
\end{equation}
In the above equations the additional factors of 2, 
  compared to their electromagnetic counterparts, 
are caused by the second-rank tensor structure of the Einstein field equations, 
 in contrast to the first-rank tensor formalism of classical electromagnetism. 
The sign differences are the result of the sign difference between the gravitational and electromagnetic forces 
  among analogous configurations of particles 
  (e.g.\ two positive charges repel one another, but two positive gravitational ``charges'' attract each other).

For clarity, let 1 and 2 denote the non-spinning and spinning bodies (the massive black hole and the orbiting DNS system), respectively.   
Thus, $m_{1} = M$ and $m_{2} = 2m$ (see \S 2). 
Let $\bm{\mathrm{r}}_{i}$ and $\bm{\mathrm{v}}_{i}$ denote the positions and velocities of bodies 1 and 2, 
  measured in the centre of momentum system. 
In the reference frame of body 2, the spin-orbit potential energy is given by  
\begin{equation}
V_{SL}^{(2)}=-2\,\bm{\mathrm{m}}_{2}\cdot\bm{\mathrm{h}}^{(2)}
  =-2\bm{\mathrm{m}}_{2}\cdot\left(\bm{\mathrm{h}}^{(1)}
   +2\frac{\bm{\mathrm{v}}}{c}\times\bm{\mathrm{g}}^{(1)}\right) \ , 
\label{GEMframeshift}
\end{equation} 
  where $\bm{\mathrm{m}}_{2}$ is the gravitational analogue of the magnetic dipole moment for body 2, 
  $\bm{\mathrm{v}}=\bm{\mathrm{v}}_{2}-\bm{\mathrm{v}}_{1}$ is the speed of body 2 
  relative to body 1, and $\bm{\mathrm{h}}^{(2)}$ is the gravito-magnetic field in the frame of body 2, 
  which transforms into fields $\bm{\mathrm{g}}^{(1)}$ and $\bm{\mathrm{h}}^{(1)}$ in the frame of body 1. 
In the rest frame of body 2, a mass current is observed as a result of the apparent motion of body 1 about body 2, 
 analogous to the apparent orbit of the proton in the electron's rest frame in the electromagnetic spin-orbit problem. 
Just as the proton's apparent current gives rise to a $\bm{\mathrm{B}}$ field in the electron's frame, 
  the apparent motion of body 1 gives rise to nonzero $\bm{\mathrm{h}}^{(2)}$. 
The additional factor of $2$ at the beginning of Eq.~\ref{GEMframeshift} 
  is due to the doubling of the $\bm{\mathrm{h}}$ field contribution in Eq.~\ref{GEMlorentz};
   the frame transformation also follows from Eq.~\ref{GEMlorentz}.
Further, $\bm{\mathrm{h}}^{(1)}=0$, 
  since the gravito-magnetic component of the field vanishes 
  after $\bm{\mathrm{h}}^{(2)}$ is transformed to the reference frame of body 1. 
That is, the gravito-magnetic field created by body 2 in the frame of body 1 is not relevant, 
  since we are interested only in transforming the fields seen in the frame of body 2. 
Equivalently, 
\begin{equation}
  V_{SL}^{(2)}=-\bm{\mathrm{p}}_{2}^{(1)}\cdot\bm{\mathrm{g}}^{(1)}\ , 
\label{GEMdipolepotential}
\end{equation}
  where 
\begin{equation}
  \bm{\mathrm{p}}_{2}^{(1)}=4\,\bm{\mathrm{m}}_{2}\times\frac{\bm{\mathrm{v}}}{c} 
\label{GEMdipole}
\end{equation}
  is the effective gravito-electric dipole moment created by the motion of $\bm{\mathrm{m}}_{2}$ in the frame of body 1. 
We note that Eq.~\ref{GEMdipole} is directly analogous to the electric dipole $\bm{\mathrm{p}}=\bm{\mathrm{v}}\times\bm{\mathrm{m}}$ 
  created by a magnetic dipole $\bm{\mathrm{m}}$ moving at velocity $\bm{\mathrm{v}}$ \citep{Fisher71}; 
  a consequence of special relativity, this effect was first predicted by~\citet{Einstein08}. 
In this case, $\bm{\mathrm{g}}^{(1)}=Gm_{1}R^{-3}\bm{\mathrm{R}}$,
where $\bm{\mathrm{R}}=\bm{\mathrm{r}}_{2}-\bm{\mathrm{r}}_{1}$
is the position of body 2 relative to body 1. Incorporating Eq.~\ref{GEMdipole}
into Eq.~\ref{GEMdipolepotential} yields  
\begin{equation}
V_{SL}^{(2)}=-\frac{4Gm_{1}}{c^{2}}\frac{\bm{\mathrm{R}}}{R^{3}}\cdot\left(\bm{\mathrm{m}}_{2}\times\bm{\mathrm{v}}\right)=\frac{4Gm_{1}}{c^{2}R^{3}}\;\bm{\mathrm{m}}_{2}\cdot\left(\bm{\mathrm{R}}\times\bm{\mathrm{v}}\right) \ . 
\label{spinorbitprelim}
\end{equation} 

A familiar result from electromagnetism is the gyromagnetic ratio $\Gamma$ 
 connecting the magnetic moment and angular momentum of a body 
 such that $\bm{\mathrm{m}}=\Gamma\bm{\mathrm{S}}$;
  for a body of mass $m$ and charge $q$, $\Gamma=q/2m$ 
  if $q/m$ is uniformly distributed. 
The gravitational analogue of $\Gamma$ can be viewed as being generated by mass currents within the body,
  due to $\bm{\mathrm{S}}$. Dividing body 2 into rings coaxial with $\bm{\mathrm{S}}_{2}$, 
  its angular momentum about its centre of mass, 
  results in $\Gamma=1/2$, that is, 
\begin{equation}
   \bm{\mathrm{m}}_{2}=\frac{1}{2}\bm{\mathrm{S}}_{2}\,,\label{gyromagnetic}
\end{equation}
   in direct analogy to the classical electromagnetic result 
  (i.e.\ an object's gravitational charge is identically its mass). 
Incorporating this result into Eq.~\ref{spinorbitprelim} yields 
\begin{equation}
    V_{SL}^{(2)}=\frac{2Gm_{1}}{c^{2}R^{3}}\;\bm{\mathrm{S}}_{2}\cdot\left(\bm{\mathrm{R}}\times\bm{\mathrm{v}}\right) \ .  
\label{spinorbitfirst}
\end{equation}

We perform a frame transformation, 
 changing from the reference frame of body 2 to the centre of momentum frame of the complete two-body system. 
In order to evaluate $V_{SL}$ in the centre of momentum frame, 
  the Thomas precession must be taken into account. 
Expressing this effect as a potential energy, we obtain 
\begin{equation}
  V_{SL}^{(\mathrm{T})}=\bm{\mathrm{S}}_{2}\cdot\bm{\omega}_{\mathrm{T},2}  \ , 
\label{Thomasfirst}
\end{equation}
  where $\bm{\omega}_{\mathrm{T,2}}=\left(\bm{\mathrm{a}}_{2}\times\bm{\mathrm{v}}_{2}\right)/2c^{2}
      =-Gm_{1}\left(\bm{\mathrm{R}}\times\bm{\mathrm{v}}_{2}\right)/2c^{2}R^{3}$,
  with $\bm{\mathrm{a}}_{2}=-Gm_{1}R^{-3}\bm{\mathrm{R}}$
  being the acceleration of body 2 in the centre of momentum frame.
This Thomas precession is a special relativistic consequence of frame transformation.  
It is a direct analogue of that obtained in changing from the frame of the electron to the centre of momentum frame 
  of the hydrogen atom in the electrodynamical problem  \citep{Jackson99}.  
In the GEM formulation effects of spacetime curvature are incorporated into the effective fields of the GEM equations. 
By substitution,
\begin{equation}
   V_{SL}^{(\mathrm{T})}=-\frac{Gm_{1}}{2c^{2}R^{3}}\;\bm{\mathrm{S}}_{2}\cdot\left(\bm{\mathrm{R}}\times\bm{\mathrm{v}}_{2}\right)\   . 
\label{Thomassecond}
\end{equation}

The orbital angular momentum of the system is $\bm{\mathrm{L}}=\bm{\mathrm{R}}\times\bm{\mathrm{P}}$
   and, to first order, $\bm{\mathrm{P}}=\mu\bm{\mathrm{v}}$,
   where $\mu=m_{1}m_{2}/\left(m_{1}+m_{2}\right)$ is the reduced mass.
Hence, Eq.~\ref{spinorbitfirst} may be re-expressed as
\begin{equation}
   V_{SL}^{(2)}=\frac{2Gm_{1}}{\mu c^{2}R^{3}}\left(\bm{\mathrm{S}}_{2}\cdot\bm{\mathrm{L}}\right)=\frac{2G}{c^{2}R^{3}}\left(1+\frac{m_{1}}{m_{2}}\right)\left(\bm{\mathrm{S}}_{2}\cdot\bm{\mathrm{L}}\right) \ . 
\label{spinorbitSL}
\end{equation}
By definition of the centre of momentum system, $m_{1}\bm{\mathrm{v}}_{1}=-m_{2}\bm{\mathrm{v}}_{2}$. 
To leading order, $\bm{\mathrm{L}}=m_{2}\bm{\mathrm{R}}\times\bm{\mathrm{v}}_{2}$.
Thus, the Thomas contribution, Eq.~\ref{Thomassecond}, may be written as
\begin{equation}
   V_{SL}^{(\mathrm{T})}=-\frac{Gm_{1}}{2m_{2}c^{2}R^{3}}\left(\bm{\mathrm{S}}_{2}\cdot\bm{\mathrm{L}}\right)\ . 
\label{ThomasSL}
\end{equation} 
The potential energy of the gravitational spin-orbit coupling in the centre of momentum frame is therefore
\begin{equation}
  V_{SL}=V_{SL}^{(2)}+V_{SL}^{(\mathrm{T})}=\frac{G}{c^{2}R^{3}}\left(2+\frac{3}{2}\frac{m_{1}}{m_{2}}\right)\left(\bm{\mathrm{S}}_{2}\cdot\bm{\mathrm{L}}\right)\ . 
\label{spinorbitfinal}
\end{equation}
For the case of the DNS system of mass $2m$, with angular momentum $\bm{\mathrm{S}}$ about its centre of mass 
  and orbiting a black hole with mass $M \gg 2m$, 
  the spin-orbit energy takes the form: 
\begin{equation}
   V_{SL}=\frac{3}{2}\frac{GM}{\left(2m\right)c^{2}R^{3}}\left(\bm{\mathrm{S}}\cdot\bm{\mathrm{L}}\right)\ . 
\label{spinorbitspecial}
\end{equation}
Note that this result has a one-to-one correspondence with that of \citet{Chan77}, 
  despite the fact that it was derived from a completely different physical setup,
  with a finite spinning mass (the DNS system) as opposed to a spinning test point particle.

\section{Orbital Dynamics} 

\subsection{Lagrangian in the orbiting reference frame}

We choose the location of the black hole as the origin of the observer's coordinates.  
Let $\mathcal{S}$ denote the distant inertial observer's frame, 
  and $\mathcal{S}'$ denote the frame rotating with angular velocity $\bm{\Omega}=\bm{\hat{\mathrm{z}}}\sqrt{GM/R^{3}}$,
  with origin at the centre of the DNS system, measured in $\mathcal{S}$ as $\bm{\mathrm{R}}$. 
Let $\bm{\mathrm{r}}$ and $\bm{\mathrm{r}}'$ represent positions measured in $\mathcal{S}$ and $\mathcal{S}'$, respectively. 
The Lagrangian of a particle of mass $m$, expressed in the coordinates of frame $\mathcal{S}'$ and coordinate time $t$, is
\begin{eqnarray}
   \mathcal{L} & =& \frac{1}{2}m\left|\dot{\bm{\mathrm{r}}}'+\bm{\Omega}\times\left(\bm{\mathrm{r}}'
     +\bm{\mathrm{R}}\right)\right|^{2}-m\Phi\left(\bm{\mathrm{r}}'\right)+\frac{GMm}{R}-V_{SL}
\nonumber \\
   & = & \frac{1}{2}m\left\{ \left|\dot{\bm{\mathrm{r}}}'\right|^{2}+2\bm{\Omega}\cdot\left[\left(\bm{\mathrm{r}}'
     +\bm{\mathrm{R}}\right)\times\dot{\bm{\mathrm{r}}}'\right]+\left[\Omega^{2}r^{2}-\left(\bm{\Omega}\cdot\bm{\mathrm{r}}\right)^{2}\right]\right\}
   -m\Phi\left(\bm{\mathrm{r}}'\right)+\frac{GMm}{R}-V_{SL} \ , 
\label{lagrangianparticle}
\end{eqnarray} 
  where $m\Phi\left(\bm{\mathrm{r}}'\right)$ is the tidal potential energy of the particle and $V_{SL}$ is the GEM spin-orbit coupling.
The dot operator denotes differentiation with respect to $t$, i.e.\ ${\dot{}} \equiv d/dt$.  
For discussions on the Lagrangian formalism in non-inertial reference frames, see e.g.\ \cite{Dallen11}.    

Here we do not consider explicitly the gravitational interaction between the two neutron stars. 
As the DNS system is represented by a spinning mass ring,  
  the gravitational interaction between two neutron stars is analogous to the constraint forces within the ring. 
As discussed in \S1 
  the distance $a/2$ of each mass element from the centre of the ring forms a holonomic constraint.  
A displacement from the configuration is an ignorable coordinate in the Lagrangian  
   and therefore there is no need to include it in the Lagrangian for the dynamical analysis of a ring.  
Nevertheless, the inclusion of such displacements in the Lagrangian would allow one 
  to calculate the tidal strain within the ring and the degree 
  to which the ring flexes for internal interaction, which is an entirely different problem from that considered here.   
For more detailed discussions on the role of holonomic constraints in the Lagrangian formulation of mechanics, 
  see \citet{Goldstein50,Fowles05}.

For the ring as a whole, of mass $2m$, 
  the Lagrangian may be found by integrating Eq.~\ref{lagrangianparticle} over the mass elements $dm$:
\begin{eqnarray}
  \mathcal{L} & = &  \frac{2GMm}{R}+\oint_{\mathrm{ring}}\frac{dm}{2}\left\{ \left|\dot{\bm{\mathrm{r}}}'\right|^{2}
  +2\bm{\Omega}\cdot\left[\left(\bm{\mathrm{r}}' 
  +\bm{\mathrm{R}}\right)\times\dot{\bm{\mathrm{r}}}'\right]+\left[\Omega^{2}r^{2}-\left(\bm{\Omega}\cdot\bm{\mathrm{r}}\right)^{2}\right]\right\} 
  -V_{SL}-\oint_{\mathrm{ring}}dm\,\Phi\left(\bm{\mathrm{r}}'\right)\  .  
\label{lagrangianint}
\end{eqnarray}
Now, 
\begin{equation}
  \oint_{\mathrm{ring}}\frac{dm}{2}\left|\dot{\bm{\mathrm{r}}}'\right|^{2}=T_{\mathrm{rot}} \ , 
\label{Trotdef}
\end{equation}
   the (rotational) kinetic energy of the ring in $\mathcal{S}'$. 
Further,
\begin{equation}
  \oint_{\mathrm{ring}}dm\,\Phi\left(\bm{\mathrm{r}}'\right)=V_{\mathrm{tidal}}  \  , 
\label{Udef}
\end{equation}
  where $V_{\mathrm{tidal}}$ is the orientational potential energy created by tidal forces. 
A further simplification of Eq.~\ref{lagrangianint} is achieved 
  by expanding 
  $\bm{\Omega}\cdot\left[\left(\bm{\mathrm{r}}'+\bm{\mathrm{R}}\right)\times\dot{\bm{\mathrm{r}}}'\right]=\bm{\Omega}\cdot    
  \left[\left(\bm{\mathrm{r}}'\times\dot{\bm{\mathrm{r}}}'\right)+\left(\bm{\mathrm{R}}\times\dot{\bm{\mathrm{r}}}'\right)\right]$.
Since $\bm{\Omega}\cdot\left(\bm{\mathrm{R}}\times\dot{\bm{\mathrm{r}}}'\right)=\dot{\bm{\mathrm{r}}}'\cdot\left(\bm{\Omega}\times\bm{\mathrm{R}}\right)$ 
  and $\bm{\Omega}\times\bm{\mathrm{R}}$ does not depend on $\bm{\mathrm{r}}'$, 
  the integral 
\begin{equation} 
\oint_{\mathrm{ring}}dm\,\bm{\Omega}\cdot\left(\bm{\mathrm{R}}\times\dot{\bm{\mathrm{r}}}'\right)=\left(\bm{\Omega}\times\bm{\mathrm{R}}\right)\cdot\oint_{\mathrm{ring}}dm\,\dot{\bm{\mathrm{r}}}'=0 \ , 
\end{equation} 
  by symmetry. 
Thus, the Lagrangian becomes 
\begin{equation}
    \mathcal{L}=T_{\mathrm{rot}}-V_{\mathrm{tidal}}-V_{SL}+\frac{2GMm}{R}
     +\oint_{\mathrm{ring}}\frac{dm}{2}\left\{ 2\bm{\Omega}\cdot\left(\bm{\mathrm{r}}'\times\dot{\bm{\mathrm{r}}}'\right)
     +\left[\Omega^{2}r^{2}-\left(\bm{\Omega}\cdot\bm{\mathrm{r}}\right)^{2}\right]\right\} \  . 
\label{lagrangianprelim}
\end{equation}

The physics is more transparent if the problem is expressed in terms of Euler angles. 
Let the 123 coordinate system be the body coordinate system moving with the orientation of the ring, 
  with the 3-axis defined as the symmetry axis. 
Define the $\mathcal{S}''$ coordinate system as follows. 
Let the $x''$-axis correspond to the line of nodes,
   that is, the intersection of the 1-2 plane (plane of the ring)
with the $x$-$y$ (equivalently, $x'$-$y'$) plane. 
Let $z''$ be aligned with the 3-axis. 
Finally, let $\alpha'$, $\beta'$, and $\gamma'$ denote the three Euler angles for the ring in frame $\mathcal{S}'$.
That is, let $\alpha'$ be the angle between the $z$- (equivalently, $z'$-) and $z''$-axes, 
  $\beta'$ the angle between the $x'$- and $x''$-axes, and $\gamma'$ the angle between the $x''$-axis and
  the 1-axis, describing rotation of the ring about its symmetry axis. 
In terms of $\bm{\hat{\mathrm{n}}}$, the normal vector to the plane of the ring, 
  with zenith and azimuth coordinates $\theta$ and $\phi$, $\alpha'=\theta$ and $\beta'=\phi-\Omega t+\pi/2$.
In terms of the Euler angles defined in the inertial frame $\mathcal{S}$,
  where $\beta$ is the angle between the line of nodes and the $x$-axis,
  $\alpha'=\alpha$, $\beta'=\beta-\Omega t$, and $\gamma'=\gamma$.
Therefore, hereafter in the analysis, primes will be dropped from $\alpha$ and $\gamma$.

The angular velocity of the spinning mass ring, which is also the angular velocity of the neutron stars, 
 in $\mathcal{S}'$ may then be expressed as 
 $\bm{\omega}'=\dot{\bm{\alpha}}+\dot{\bm{\beta}}'+\dot{\bm{\gamma}}$.
In terms of the $123$ coordinate system,  
\begin{eqnarray}
\omega_{1}' & = & \dot{\beta}'\sin\alpha\sin\gamma+\dot{\alpha}\cos\gamma  \ ;   \nonumber \\
\omega_{2}' & = & \dot{\beta}'\sin\alpha\cos\gamma-\dot{\alpha}\sin\gamma \ ;  \nonumber \\
\omega_{3}' & = & \dot{\beta}'\cos\alpha+\dot{\gamma}  \ .   \label{omega123}
\end{eqnarray}
The kinetic energy $T_{\mathrm{rot}}$ is given in $\mathcal{S}'$
   by $\left(1/2\right)\sum_{i=1}^{3}I_{i}\omega_{i}'^{2}$, where $I_{1}=I_{2}=\left(1/2\right)\left(2m\right)\left(a/2\right)^{2}=ma^{2}/4$
   and $I_{3}=\left(2m\right)\left(a/2\right)^{2}=ma^{2}/2$. 
Thus, 
\begin{eqnarray}
T_{\mathrm{rot}} & = & \frac{1}{8}ma^{2}\bigl(\dot{\beta}'^{2}\sin^{2}\alpha\sin^{2}\gamma+\dot{\alpha}^{2}\cos^{2}\gamma+2\dot{\alpha}\dot{\beta}'\sin\alpha\sin\gamma\cos\gamma 
 +\dot{\beta}'^{2}\sin^{2}\alpha\cos^{2}\gamma+\dot{\alpha}^{2}\sin^{2}\gamma-2\dot{\alpha}\dot{\beta}'\sin\alpha\sin\gamma\cos\gamma
  \nonumber \\
 &  & \hspace{1.0cm}  +2\dot{\beta}'^{2}\cos^{2}\alpha+2\dot{\gamma}^{2}+4\dot{\beta}'\dot{\gamma}\cos\alpha\bigr) \nonumber \\
 & = & \frac{1}{8}ma^{2}\left[\dot{\beta}'^{2}\left(1+\cos^{2}\alpha\right)+\dot{\alpha}^{2}+2\dot{\gamma}^{2}+4\dot{\beta}'\dot{\gamma}\cos\alpha\right] \  .
 \label{Trot}
\end{eqnarray} 

The non-inertial terms in the Lagrangian can be divided into a Coriolis term 
\begin{equation}
   T_{\mathrm{Cor}}=\oint_{\mathrm{ring}}dm\,\bm{\Omega}\cdot\left(\bm{\mathrm{r}}'\times\dot{\bm{\mathrm{r}}}'\right)
\label{Tcorprelim}
\end{equation}
and a centrifugal term 
\begin{equation}
   T_{\mathrm{cen}}=\frac{1}{2}\oint_{\mathrm{ring}}dm\left[\Omega^{2}r^{2}-\left(\bm{\Omega}\cdot\bm{\mathrm{r}}\right)^{2}\right]\  . 
\label{Tcenprelim}
\end{equation}  
The Coriolis term may be simplified by expressing $\dot{\bm{\mathrm{r}}}'$ as $\bm{\omega}'\times\bm{\mathrm{r}}'$. 
We may rewrite $\bm{\Omega}\cdot\left(\bm{\mathrm{r}}'\times\dot{\bm{\mathrm{r}}}'\right)$ as  
\begin{equation} 
 \bm{\Omega}\cdot\left[\bm{\mathrm{r}}'\times\left(\bm{\omega}'\times\bm{\mathrm{r}}'\right)\right]=\bm{\Omega}\cdot\left[\bm{\omega}'\left(\bm{\mathrm{r}}'\cdot\bm{\mathrm{r}}'\right)-\bm{\mathrm{r}}'\left(\bm{\mathrm{r}}'\cdot\bm{\omega}'\right)\right]=\bm{\Omega}\cdot\bm{\omega}'r'^{2}-\left(\bm{\Omega}\cdot\bm{\mathrm{r}}'\right)\left(\bm{\omega}'\cdot\bm{\mathrm{r}}'\right)  \ . 
 \end{equation}   
For the rigid mass ring, $r'=a/2$ is a constant. 
In terms of the $\mathcal{S}''$ coordinate system, 
  $\bm{\Omega}=\Omega\left(\bm{\hat{\mathrm{y}}}''\sin\alpha+\bm{\hat{\mathrm{z}}}''\cos\alpha\right)$
  and $\bm{\omega}'=\bm{\hat{\mathrm{x}}}''\dot{\alpha}+\bm{\hat{\mathrm{y}}}''\dot{\beta}'\sin\alpha
   +\bm{\hat{\mathrm{z}}}''\left(\dot{\beta}'\cos\alpha+\dot{\gamma}\right)$.
For a mass element $dm$ along the ring,  specified by the Euler angle
  $\gamma$, $\bm{\mathrm{r}}'=r'\left(\bm{\hat{\mathrm{x}}}''\cos\gamma+\bm{\hat{\mathrm{y}}}''\sin\gamma\right)$.
Thus, $\bm{\Omega}\cdot\bm{\mathrm{r}}'=\Omega\, r'\sin\alpha\sin\gamma$,
   $\bm{\omega}'\cdot\bm{\mathrm{r}}'=r'\left(\dot{\alpha}\cos\gamma+\dot{\beta}'\sin\alpha\sin\gamma\right)$, and 
  $\bm{\Omega}\cdot\bm{\omega}'=\Omega\left(\dot{\beta}'\sin^{2}\alpha+\dot{\beta}'\cos^{2}\alpha+\dot{\gamma}\cos\alpha\right)=\Omega\left(\dot{\beta}'
    +\dot{\gamma}\cos\alpha\right)$.
Therefore, expressing $dm$ as $2m\, d\gamma/2\pi$, 
\begin{eqnarray} 
   T_{\mathrm{Cor}} & = &  
   2m\left(\frac{a}{2}\right)^{2}\bm{\Omega}\cdot\bm{\omega}'-\frac{2m}{2\pi}\left(\frac{a}{2}\right)^{2}\Omega 
    \int_{0}^{2\pi}d\gamma\left(\sin\alpha\sin\gamma\right) 
   \left(\dot{\alpha}\cos\gamma+\dot{\beta}'\sin\alpha\sin\gamma\right)  
    \nonumber \\
  & = & \frac{1}{2}ma^{2}\Omega\left[\dot{\beta}'\left(1-\frac{1}{2}\sin^{2}\alpha\right)+\dot{\gamma}\cos\alpha\right] \ . 
\label{Tcor}
\end{eqnarray} 
Further, the centrifugal term may be simplified by expansion. 
Since $\bm{\mathrm{r}}=\bm{\mathrm{R}}+\bm{\mathrm{r}}'$,
we have
\begin{eqnarray}
\Omega^{2}r^{2}-\left(\bm{\Omega}\cdot\bm{\mathrm{r}}\right)^{2} 
  & = & \Omega^{2}\left(\bm{\mathrm{R}}+\bm{\mathrm{r}}'\right)\cdot\left(\bm{\mathrm{R}}+\bm{\mathrm{r}}'\right)
    -\left[\bm{\Omega}\cdot\left(\bm{\mathrm{R}}+\bm{\mathrm{r}}'\right)\right]^{2} 
    \nonumber \\
 & = & \Omega^{2}\left(R^{2}+r'^{2}+2\bm{\mathrm{R}}\cdot\bm{\mathrm{r}}'\right)-\left(\bm{\Omega}\cdot\bm{\mathrm{R}}\right)^{2}
   -\left(\bm{\Omega}\cdot\bm{\mathrm{r}}'\right)^{2}-2\left(\bm{\Omega}\cdot\bm{\mathrm{R}}\right)\left(\bm{\Omega}\cdot\bm{\mathrm{r}}'\right) \ . 
\label{Omegar}
\end{eqnarray}
Now, $\bm{\Omega}\perp\bm{\mathrm{R}}$ and $\oint_{\mathrm{ring}}dm\,\bm{\mathrm{R}}\cdot\bm{\mathrm{r}}'=\bm{\mathrm{R}}\cdot\oint_{\mathrm{ring}}dm\,\bm{\mathrm{r}}'=0$, by symmetry. 
Thus, with $\bm{\Omega}\cdot\bm{\mathrm{r}}'=\Omega r'\sin\alpha\sin\gamma$, as shown previously, 
  the centrifugal contribution to the Lagrangian is  
\begin{eqnarray}
   T_{\mathrm{cen}} & = & \frac{1}{2}\oint_{\mathrm{ring}}dm\left[\Omega^{2}r'^{2}-\left(\bm{\Omega}\cdot\bm{\mathrm{r}}'\right)^{2}\right]
    +m\Omega^{2}R^{2} 
    \nonumber \\
    & = & \frac{1}{2}\left(2m\right)\left(\frac{a}{2}\right)^{2}\Omega^{2}+m\Omega^{2}R^{2}-\frac{1}{2}\left(\frac{2m}{2\pi}\right) 
      \left(\frac{a}{2}\right)^{2}\Omega^{2}\sin^{2}\alpha\int_{0}^{2\pi}d\gamma\sin^{2}\gamma 
    \nonumber \\
   & = & \frac{1}{4}ma^{2}\Omega^{2}\left(1-\frac{1}{2}\sin^{2}\alpha\right)+m\Omega^{2}R^{2} \ . 
\label{Tcen}
\end{eqnarray}
Upon inspection of Eq.~\ref{Tcen}, one can see that in fact the centrifugal term can be interpreted as a potential energy in the Lagrangian,
   with $V_{\mathrm{cen}}=-T_{\mathrm{cen}}$. 

The tidal orientational potential energy was derived in Eq.~\ref{tidalpotentialint} in terms of the direction of $\bm{\hat{\mathrm{n}}}$.  
It becomes, in terms of the Euler angles, 
\begin{equation}
   V_{\mathrm{tidal}}=\frac{GMm}{R}\left(\frac{a}{2R}\right)^{2}\left\{ 1-\frac{3}{2}\left[\cos^{2}\alpha\sin^{2}\beta'+\cos^{2}\beta'\right]\right\}\  . 
\label{Vtidal}
\end{equation}

Finally, the relativistic spin-orbit potential energy must be expressed in terms of the Euler angles. 
The orbital angular momentum is $\bm{\mathrm{L}}=2mR^{2}\bm{\Omega}$
  and the angular momentum of the ring is $\bm{\mathrm{S}}=\mathbb{I}\bm{\omega}$,
  where $\mathbb{I}$ is the ring's moment of inertia tensor and $\bm{\omega}$
   is the angular velocity of the ring defined in $\mathcal{S}$. 
With $\bm{\hat{\mathrm{e}}}_{i}$ denoting the orthonormal basis vectors of the 123 coordinate system, 
     $\bm{\omega}=\bm{\hat{\mathrm{e}}}_{1}\left(\dot{\beta}\sin\alpha\sin\gamma+\dot{\alpha}\cos\gamma\right)
        +\bm{\hat{\mathrm{e}}}_{2}\left(\dot{\beta}\sin\alpha\cos\gamma-\dot{\alpha}\sin\gamma\right)
        +\bm{\hat{\mathrm{e}}}_{3}\left(\dot{\beta}\cos\alpha+\dot{\gamma}\right)$.
Note that here we use $\dot{\beta}$, not $\dot{\beta}'$, 
  since it is the spin observed in the inertial system that determines $V_{SL}$.
Working in the $123$ coordinate system, we have:
\begin{eqnarray}
     \bm{\mathrm{S}} & = & \frac{1}{2}\left(2m\right)\left(\frac{a}{2}\right)^{2}\left[\left(\dot{\beta}\sin\alpha\sin\gamma+\dot{\alpha}\cos\gamma\right) 
     \bm{\hat{\mathrm{e}}}_{1} +\left(\dot{\beta}\sin\alpha\cos\gamma-\dot{\alpha}\sin\gamma\right)\bm{\hat{\mathrm{e}}}_{2} 
     +2\left(\dot{\beta}\cos\alpha+\dot{\gamma}\right)\bm{\hat{\mathrm{e}}}_{3}\right] 
\label{S123}
\end{eqnarray}
and 
\begin{equation}
   \bm{\Omega}=\Omega\left(\bm{\hat{\mathrm{e}}}_{1}\sin\alpha\sin\gamma
   +\bm{\hat{\mathrm{e}}}_{2}\sin\alpha\cos\gamma+\bm{\hat{\mathrm{e}}}_{3}\cos\alpha\right) \ , 
\label{Omega123}
\end{equation}
whence 
\begin{eqnarray}
   \bm{\mathrm{S}}\cdot\bm{\mathrm{L}} 
    & = & \frac{1}{2}\left(2m\right)^{2}\left(\frac{a}{2}\right)^{2}R^{2}\Omega\biggl[\dot{\beta}\sin^{2}\alpha\sin^{2}\gamma
    +\dot{\alpha}\sin\alpha\sin\gamma\cos\gamma   
    +\dot{\beta}\sin^{2}\alpha\cos^{2}\gamma-\dot{\alpha}\sin\alpha\sin\gamma\cos\gamma+2\dot{\beta}\cos^{2}\alpha     
    +2\dot{\gamma}\cos\alpha\biggr] 
 \nonumber \\
    &  = & \frac{1}{2}m^{2}a^{2}R^{2}\Omega\left[\dot{\beta}\left(1+\cos^{2}\alpha\right)+2\dot{\gamma}\cos\alpha\right] \ .
\label{SdotL}
\end{eqnarray}  
Thus,
\begin{eqnarray}
  V_{SL} & = & \frac{3}{2}\frac{GM}{\left(2m\right)c^{2}R^{3}}\left(\bm{\mathrm{S}}\cdot\bm{\mathrm{L}}\right) 
  \nonumber \\
 & = & \frac{3}{8}\frac{GMm}{R}\frac{a^{2}\Omega}{c^{2}}\left[\left(\dot{\beta}'+\Omega\right)
   \left(1+\cos^{2}\alpha\right)+2\dot{\gamma}\cos\alpha\right] \ .
\label{VSL}
\end{eqnarray} 

Finally, the complete Lagrangian for a DNS system of mass $2m$ and diameter $a$ in circular orbit 
  at distance $R\gg a$ from a black hole of mass $M\gg2m$ may be written:
\begin{eqnarray}
\mathcal{L} & \hspace*{-0.25cm} =  \frac{1}{8}ma^{2}\left[\dot{\beta}'^{2}\left(1+\cos^{2}\alpha\right)+\dot{\alpha}^{2}+2\dot{\gamma}^{2}+4\dot{\beta}'\dot{\gamma}\cos\alpha\right] & \left[T_{\mathrm{rot}}\right]
  \nonumber \\
 &   \hspace*{-0.53cm}+\frac{1}{2}ma^{2}\Omega\left[\dot{\beta}'\left(1-\frac{1}{2}\sin^{2}\alpha\right)+\dot{\gamma}\cos\alpha\right] 
  &   \left[T_{\mathrm{Cor}}\right]
  \nonumber \\
 &   \hspace*{-2.39cm}+\frac{1}{4}ma^{2}\Omega^{2}\left(1-\frac{1}{2}\sin^{2}\alpha\right) & \left[V_{\mathrm{cen}}\right]\nonumber \\
 &   \hspace*{0.8cm}-\frac{GMm}{R}\left(\frac{a}{2R}\right)^{2}\left[1-\frac{3}{2}\big(\cos^{2}\alpha\sin^{2}\beta'+\cos^{2}\beta'\big)\right] & \left[V_{\mathrm{tidal}}\right]\nonumber \\
 &   \hspace*{0.92cm}-\frac{3}{8}\frac{GMm}{R}\frac{a^{2}\Omega}{c^{2}}\left[\big(\dot{\beta}'+\Omega\big)\big(1+\cos^{2}\alpha\big)+2\dot{\gamma}\cos\alpha\right] & \left[V_{SL}\right]\nonumber \\
 &   \hspace*{-3.16cm}+m\Omega^{2}R^{2}+\frac{2GMm}{R}  \  .  & \left[\mathrm{constants}\right] 
 \label{BigLagrangianprime}
\end{eqnarray} 
In terms of $\Omega=\sqrt{GM/R^{3}}$ where possible, 
  we may express the Lagrangian as
\begin{eqnarray}
\mathcal{L} & \hspace*{-0.5cm}=\frac{1}{8}ma^{2}\left[\dot{\beta}'^{2}\left(1+\cos^{2}\alpha\right)+\dot{\alpha}^{2}+2\dot{\gamma}^{2}+4\dot{\beta}'\dot{\gamma}\cos\alpha\right] & \left[T_{\mathrm{rot}}\right] 
   \nonumber \\
 & \hspace*{-0.5cm}+\frac{1}{2}ma^{2}\Omega\left[\dot{\beta}'\left(1-\frac{1}{2}\sin^{2}\alpha\right)+\dot{\gamma}\cos\alpha\right] & \left[T_{\mathrm{Cor}}\right] 
    \nonumber \\
 & \hspace*{-2.35cm}+\frac{1}{4}ma^{2}\Omega^{2}\left(1-\frac{1}{2}\sin^{2}\alpha\right) & \left[V_{\mathrm{cen}}\right] 
   \nonumber \\
 & \hspace*{0.22cm}-\frac{1}{4}ma^{2}\Omega^{2}\left[1-\frac{3}{2}\left(\cos^{2}\alpha\sin^{2}\beta'+\cos^{2}\beta'\right)\right] & \left[V_{\mathrm{tidal}}\right] 
   \nonumber \\
 & \hspace*{0.91cm}-\frac{3}{8}m\Omega^{3}\frac{a^{2}R^{2}}{c^{2}}\left[\left(\dot{\beta}'+\Omega\right)\left(1+\cos^{2}\alpha\right)+2\dot{\gamma}\cos\alpha\right] & \left[V_{SL}\right] 
   \nonumber \\
 & \hspace*{-4.2cm}+3m\Omega^{2}R^{2}\,. & \left[\mathrm{constants}\right] 
\label{BigLagrangianpg}
\end{eqnarray}
Grouping like terms, applying trigonometric identities, 
  and rewriting Eq.~\ref{BigLagrangianpg} in terms of $\beta=\beta'+\Omega t$,
  that is, in the Euler angles of the inertial frame, we obtain 
\begin{eqnarray}
\mathcal{L} & \hspace*{-1.83cm} 
  = \frac{1}{8}ma^{2}\left[\dot{\beta}^{2}\left(1+\cos^{2}\alpha\right)+\dot{\alpha}^{2}+2\dot{\gamma}^{2}+4\dot{\beta}\dot{\gamma}\cos\alpha\right] 
  & \left[T_{\mathrm{rot}},T_{\mathrm{Cor}},V_{\mathrm{cen}}\right] 
\nonumber \\
 & \hspace*{1,0cm}-\frac{1}{4}ma^{2}\Omega^{2}\left\{ 1-\frac{3}{2}\left[\cos^{2}\alpha\sin^{2}\left(\beta-\Omega t\right)+\cos^{2}\left(\beta-\Omega t\right)\right]\right\}  & \left[V_{\mathrm{tidal}}\right] 
 \nonumber \\
 & \hspace*{-1.27cm}-\frac{3}{8}m\Omega^{3}\frac{a^{2}R^{2}}{c^{2}}\left[\dot{\beta}\left(1+\cos^{2}\alpha\right)+2\dot{\gamma}\cos\alpha\right] & \left[V_{SL}\right]
 \nonumber \\
 & \hspace*{-5.18cm}+3m\Omega^{2}R^{2}\ . & \left[\mathrm{constants}\right] 
 \label{BigLagrangianS}
\end{eqnarray}
$T_{\mathrm{rot}}$, $T_{\mathrm{Cor}}$, and $V_{\mathrm{cen}}$ combine to yield the rotational kinetic energy seen by an observer 
  in the inertial frame. 
Through this analysis, we have shown explicitly that the Lagrangian can be correctly formulated in the inertial frame, 
  provided that the tidal potential seen in the non-inertial frame undergoes time-dependent coordinate rotation at rate $\Omega$. 
Coriolis and centrifugal forces do not otherwise affect the results.

\subsection{Equations of motion}

The Euler-Lagrange equations governing the dynamics of the DNS system are  
\begin{equation}
\frac{d}{dt}\left(\frac{\partial\mathcal{L}}{\partial\dot{\psi}}\right)=\frac{\partial\mathcal{L}}{\partial\psi}\,,\label{LagrangesEqn}
\end{equation}
where $\psi=\alpha,\;\beta,\;\gamma$ and, 
 as previously, the dot operator signifies differentiation with respect to coordinate time, $\dot{}\equiv {d}/{dt}$, 
 that is, the time measured by a distant inertial observer. 
Proper time, experienced in the reference frame of a particle, is defined differentially 
 as $c\, d\tau=\sqrt{dx_{\mu}dx^{\mu}}=\sqrt{g_{\mu\nu}dx^{\mu}dx^{\nu}}$, where $g_{\mu\nu}$ is the spacetime metric tensor. 
The Schwarzschild metric, expressed in spherical coordinates, is
\begin{equation}
   c^{2}d\tau^{2}=\left(1-\frac{r_{\mathrm{s}}}{r}\right)c^{2}dt^{2}-\left(1-\frac{r_{\mathrm{s}}}{r}\right)^{-1}dr^{2}-r^{2}d\theta^{2}-r^{2}\sin^{2}\theta d\phi^{2}\,,\label{Schmetric}
\end{equation}
where $r_{\mathrm{s}}=2GM/c^{2}$ is the Schwarzschild radius. 
For particles at rest at some radius $R$ in the Schwarzschild metric 
 ($dr=d\theta=d\phi=0$), $d\tau=dt\sqrt{1-2GM/c^{2}R}$. 
For particles in circular orbit at that radius, $dr=d\theta=0$ and $\theta=\pi/2$, but $d\phi=\Omega dt$. 
Here, $\Omega$ is the usual Keplerian angular velocity, $\sqrt{GM/R^{3}}$, 
 which is the exact general relativistic result in coordinate time, regardless of $R$  \citep{Misner77}. 
Thus, for the orbiting particle,  
\begin{equation}
  c^{2}d\tau^{2}=\left(1-\frac{2GM}{Rc^{2}}\right)c^{2}dt^{2}-R^{2}\frac{GM}{R^{3}}dt^{2}=\left(1-\frac{3GM}{Rc^{2}}\right)c^{2}dt^{2} \  . 
\label{propertimemetric}
\end{equation}
It follows that  
\begin{equation}
   d\tau=dt\sqrt{1-\frac{3GM}{Rc^{2}}}  \   . 
\label{propertime}
\end{equation}

Taking the relevant derivatives of Eq.~\ref{BigLagrangianS} and rearranging, 
 we obtain the equations of motion for the DNS system: 
\begin{eqnarray}
 0 & = & \ddot{\alpha}+\left(\dot{\beta}^{2}-3\Omega^{3}\frac{R^{2}}{c^{2}}\dot{\beta}\right)\sin\alpha\cos\alpha 
   +\left(2\dot{\beta}-3\Omega^{3}\frac{R^{2}}{c^{2}}\right)\dot{\gamma}\sin\alpha+3\Omega^{2}\sin^{2}\left(\beta-\Omega t\right)\sin\alpha\cos\alpha \  ;
\label{EoMa} \\
 0 & = & \ddot{\beta}\left(1+\cos^{2}\alpha\right)-\left(2\dot{\beta}-3\Omega^{3}\frac{R^{2}}{c^{2}}\right)\dot{\alpha}\sin\alpha\cos\alpha 
  -2\dot{\alpha}\dot{\gamma}\sin\alpha+2\ddot{\gamma}\cos\alpha+3\Omega^{2}\sin^{2}\alpha\sin\left(\beta-\Omega t\right)\cos\left(\beta-\Omega t\right) \  ; 
\label{EoMb} \\
  0 & = & 2\ddot{\gamma}+2\ddot{\beta}\cos\alpha-\left(2\dot{\beta}-3\Omega^{3}\frac{R^{2}}{c^{2}}\right)\dot{\alpha}\sin\alpha \  . 
\label{EoMc}
\end{eqnarray}
Defining a new parameter $\dot{\xi}\equiv2\dot{\beta}-3\Omega^{3}R^{2}c^{-2}$,
Eqs.~\ref{EoMa} through~\ref{EoMc} can be expressed more compactly: 
\begin{eqnarray}
0 & = & \ddot{\alpha}+\left(\dot{\xi}-\dot{\beta}\right)\dot{\beta}\sin\alpha\cos\alpha+\dot{\xi}\dot{\gamma}\sin\alpha+3\Omega^{2}\sin^{2}\left(\beta-\Omega t\right)\sin\alpha\cos\alpha \ ; 
\label{EoMaXi} \\
0 & = & \ddot{\beta}\left(1+\cos^{2}\alpha\right)-\dot{\xi}\dot{\alpha}\sin\alpha\cos\alpha-2\dot{\alpha}\dot{\gamma}\sin\alpha 
 +2\ddot{\gamma}\cos\alpha+3\Omega^{2}\sin^{2}\alpha\sin\left(\beta-\Omega t\right)\cos\left(\beta-\Omega t\right)
 \  ;     
 \label{EoMbXi} \\ 
0 & = & 2\ddot{\gamma}+2\ddot{\beta}\cos\alpha-\dot{\xi}\dot{\alpha}\sin\alpha \  . 
\label{EoMcXi}
\end{eqnarray}

\section{Results and Discussions}

\subsection{Demonstration of restricted oscillations}

To understand some of the dynamical properties of this system, 
  it is instructive to consider the special cases in which the tidal effects are singled out.  
If we let the mass ring (the DNS system) be held at fixed location in space by a force,   
  the time dependence of Eq.~\ref{BigLagrangianS} is suppressed.   
Without loss of generality, we fix the centre of mass of the ring to lie along the positive $x$-axis, 
  so that the orientational potential energy, now time-independent, may be expressed in terms
  of the coordinates of $\bm{\hat{\mathrm{n}}}$:
\begin{equation}
    V_{\mathrm{tidal}}=\frac{GMm}{R}\left(\frac{a}{2R}\right)^{2}\left[1-\frac{3}{2}\left(\cos^{2}\theta\cos^{2}\phi+\sin^{2}\phi\right)\right] \ . 
\label{Vtidalstat}
\end{equation}
The behaviour of these special, restricted cases will be useful in understanding the more complex nature of solutions 
  to the general case of the orbiting, unconstrained ring. 
The Lagrangian for this restricted system is 
\begin{eqnarray}
    \mathcal{L}_{\mathrm{stat}} & = & \frac{1}{8}ma^{2}\left[\dot{\theta}^{2}+\left(1+\cos^{2}\theta\right)\dot{\phi}^{2}\right] 
    -\frac{GMm}{R}\left(\frac{a}{2R}\right)^{2}\left[1-\frac{3}{2}\left(\cos^{2}\theta\cos^{2}\phi+\sin^{2}\phi\right)\right] \  ,  
\label{restrictedLagrangian}
\end{eqnarray}  
  which implies that the oscillations have no explicit dependence on the orbital parameters 
  of the neutron stars in the DNS system, in particular, the orbital separation $a$. 

We consider two cases to demonstrate the restricted oscillations arising from such situations. 
In the first case, we  let $\phi$ be fixed and let $\theta$ be free; in the second case let $\theta$ be fixed and let $\phi$ be free. 
In the calculations, the system parameters are 
  $M=1.5\times10^{7}\; M_{\odot}$ (black hole), 
  $m=1.5\; M_{\odot}$ (neutron star), 
  $R=25\ r_{\mathrm{s}}$, where the Schwarzschild radius of the black hole $r_{\mathrm{s}}=2GM/c^{2}=4.4\times10^{12}\,\mathrm{cm}$, and
  $\Omega=\sqrt{GM/R^{3}}=3.8\times10^{-5}\,\mathrm{rad\, s^{-1}}$.  

\subsubsection{Case 1: fixed $\phi$, free $\theta$} 
 
Lagrange's equation yields 
\begin{equation}
\ddot{\theta}+3\Omega^{2}\cos^{2}\phi\sin\theta\cos\theta=0 \ . 
\label{Thetaspecial}
\end{equation}
By inspection of Eq.~\ref{Thetaspecial}, 
  one can see that $\theta$ has stable equilibria at integer multiples of $\pi$ and unstable equilibria 
  at odd half-integer multiples of $\pi$.  
It is easy to visualize in the $\phi=0$ case: 
  tidal forces will try to pull the mass ring into the plane defined by $\mathbf{R}$ and $\mathbf{R}\times\mathbf{L}$, 
  that is, the $x$-$y$ plane. 
For small $\theta$, $\sin\theta\cos\theta\approx\theta$ and oscillation about stable equilibrium will occur with angular frequency 
\begin{equation}
   \omega_{\theta,0}
   =\Omega\sqrt{3}\left|\cos\phi\right| \ , 
\label{omega0theta}
\end{equation}
  that is, $\theta=\theta_{0}\cos\omega_{\theta,0}t$.

This estimate may be improved by the method of successive approximations,
   using the next term in the Taylor expansion, $\cos\theta\sin\theta\approx\theta-\left(2\theta^{2}/3\right)$,
   and the result obtained in the first approximation. 
This process results in a slight lengthening of the period of oscillation and the appearance of higher harmonics. 
At next-to-leading order, 
\begin{equation}
    \omega_{\theta}=\omega_{\theta,0}\sqrt{1-\frac{1}{2}\theta_{0}^{2}}
\label{omegatheta}
\end{equation}
and
\begin{equation}
    \theta=\theta_{0}\cos\omega_{\theta}t+\left(27\theta_{0}^{2}-48\right)^{-1}\theta_{0}^{3}\cos3\omega_{\theta}t \ . 
\label{thetaharm}
\end{equation}

\begin{figure} 
\vspace*{0.25cm}
\noindent 
\begin{centering} 
\includegraphics[width=6.5cm]{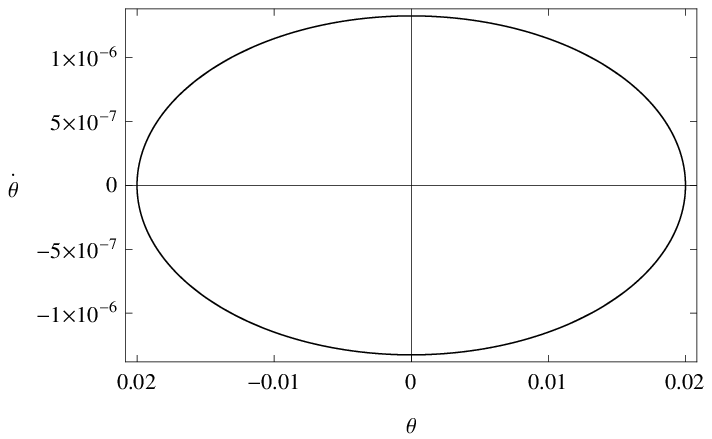}\hspace*{0.5cm}
\includegraphics[width=6.5cm]{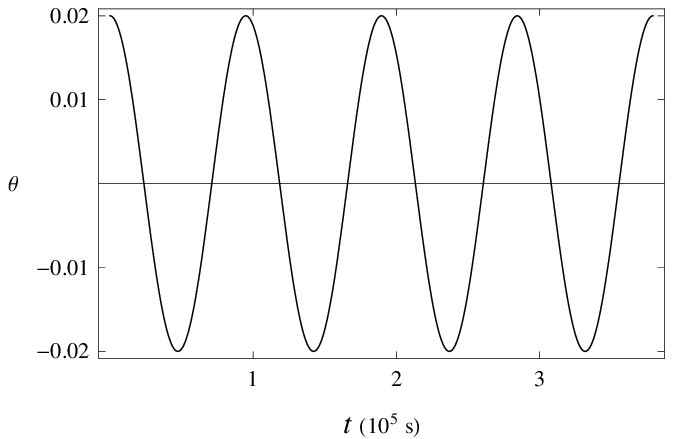}
\par
\end{centering} 
\vspace*{0.25cm} 
\par
\noindent 
\begin{centering}
\includegraphics[width=6.5cm]{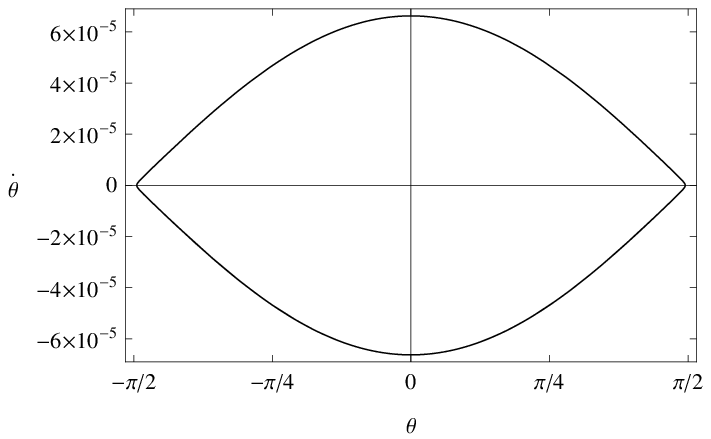}\hspace*{0.54cm}
\includegraphics[width=6.5cm]{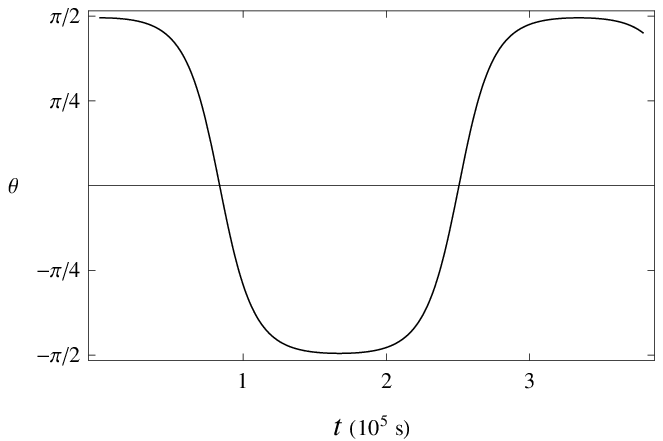}
\par
\end{centering} 
\vspace*{0.3cm}
\caption{ Phase space and time-dependent behaviour in the zenith angle $\theta$ 
  for oscillation of the fixed, $\phi$-restricted, non-rotating ring. 
For small oscillations, the behaviour is simply harmonic, 
  while for oscillations with amplitude approaching $\pi/2$, 
  lingering behaviour occurs as a result of the unstable equilibrium.
In the calculation  
  we use $M=1.5\times10^{7}\; M_{\odot}$ (black hole), 
  $m=1.5\; M_{\odot}$ (neutron stars), $R=25r_{\mathrm{s}}$,
  where $r_{\mathrm{s}}=2GM/c^{2}=4.4\times10^{12}\,\mathrm{cm}$, and
  $\Omega=\sqrt{GM/R^{3}}=3.8\times10^{-5}\,\mathrm{rad\, s^{-1}}$.
The units of $\dot{\theta}$ are $\mathrm{rad\, s^{-1}}$.  \
} 
\label{phasefig} 
\end{figure}

In phase space ($\theta-\dot{\theta}$), when $\theta_{0}$ is small,
  the path traces an ellipse, as required of simple harmonic oscillation.
However, as $\theta_{0}$ approaches $\pi/2$,  
  $\bm{\hat{\mathrm{n}}}$ tends to hang longer near this point of unstable equilibrium, 
  making the time-dependent behaviour of $\theta$ approach a square-like wave.
Likewise, the phase plot becomes somewhat pinched 
  and the phase trajectory approaches a homoclinic orbit  
  as the initial condition approaches unstable equilibrium 
  (see Fig.~\ref{phasefig}). 


\subsubsection{Case II: fixed $\theta$, free $\phi$}  

The application of Lagrange's equation to Eq.\ \ref{restrictedLagrangian} yields
\begin{equation}
  \left(1+\cos^{2}\theta\right)\ddot{\phi}- 3\Omega^{2} \left(1-\cos^{2}\theta\right)\sin\phi\cos\phi=0 \ . 
\label{Phispecial}
\end{equation}
As one can see, Eq.~\ref{Phispecial} possesses stable equilibria at odd half-integer multiples of $\pi$ and unstable equilibria 
   at integer multiples of $\pi$. 
As in the case of $\theta$, small oscillations in $\phi$ about the stable equilibria may be treated as simply harmonic,
   with natural frequency
\begin{equation}
   \omega_{\phi,0}
     =\Omega\sqrt{3}\sqrt{\frac{1-\cos^{2}\theta}{1+\cos^{2}\theta}} \ . 
\label{omega0phi}
\end{equation}

This case, too, may be corrected through successive approximations,
   with identical results, for $\phi$ near $\pi/2$:
\begin{equation}
  \omega_{\phi}=\omega_{\phi,0}\sqrt{1-\frac{1}{2}\phi_{0}^{2}} 
\label{omegaphi}
\end{equation}
and
\begin{equation}
  \phi=\frac{\pi}{2}+\phi_{0}\cos\omega_{\phi}t+\left(27\phi_{0}^{2}-48\right)^{-1}\phi_{0}^{3}\cos3\omega_{\phi}t \ , 
\label{phiharm}
\end{equation}
   where $\phi_{0}\equiv\phi\left(t=0\right)-\pi/2$. 
The behaviour of $\phi$ in time and phase space is analogous to that shown in Fig.~\ref{phasefig},
    but shifted by $\pi/2$.

Thus, the natural oscillatory frequency of the mass ring 
  due to tidal forces scales with the orbital frequency $\Omega$. 
This relationship causes the orbit to function as a driving frequency for the case of the mass ring
   in orbit about the central mass.  
In the reference frame of the ring, the effective tidal potential of Fig.~\ref{tidalfig} is spinning  at frequency $\Omega$. 
However, this resonant behaviour is highly dependent upon initial conditions and the spin of the mass ring.

\subsection{General situation}

Finally, we demonstrate the nutational and precessional behaviour of the general case. 
A numerical solution of Eqs.~\ref{EoMa} through~\ref{EoMc} is calculated for the same parameters as previously:
   $M=1.5\times10^{7}\; M_{\odot}$ (massive black hole), $m=1.5\; M_{\odot}$ (neutron star), 
   $R=25~r_{\mathrm{s}}$, where $r_{\mathrm{s}}=2GM/c^{2}=4.4\times10^{12}\,\mathrm{cm}$, 
   and $\Omega=\sqrt{GM/R^{3}}=3.8\times10^{-5}\,\mathrm{rad\, s^{-1}}$,
   with the initial conditions $\alpha\left(t=0\right)=0.02\,\mathrm{rad}$,
   $\beta\left(t=0\right)=\pi/2$, $\gamma\left(t=0\right)=0$, $\dot{\alpha}\left(t=0\right)=\dot{\beta}\left(t=0\right)=0$,
   and $\dot{\gamma}\left(t=0\right)=\sqrt{2Gm/a^{3}}=3.8\times10^{-4}\,\mathrm{rad\, s^{-1}}$,
   where $a=2\; R_{\odot}$. 
These conditions were chosen to illustrate the effect of a small angular perturbation on the orbiting system;
   $\dot{\gamma}\left(t=0\right)$ was chosen to represent the Keplerian angular speed of the analogous DNS system 
   for the mass and size of the ring. 

\begin{figure}
\noindent 
\vspace*{0.4cm}  \\  
\begin{centering} 
\noindent
\includegraphics[width=6.5cm]{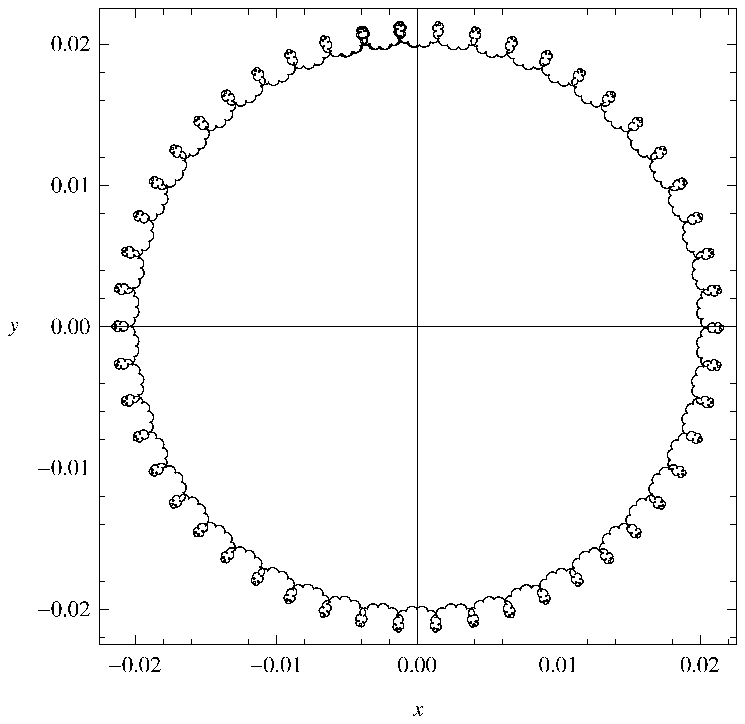} \hspace*{1cm}  
\includegraphics[width=6.5cm]{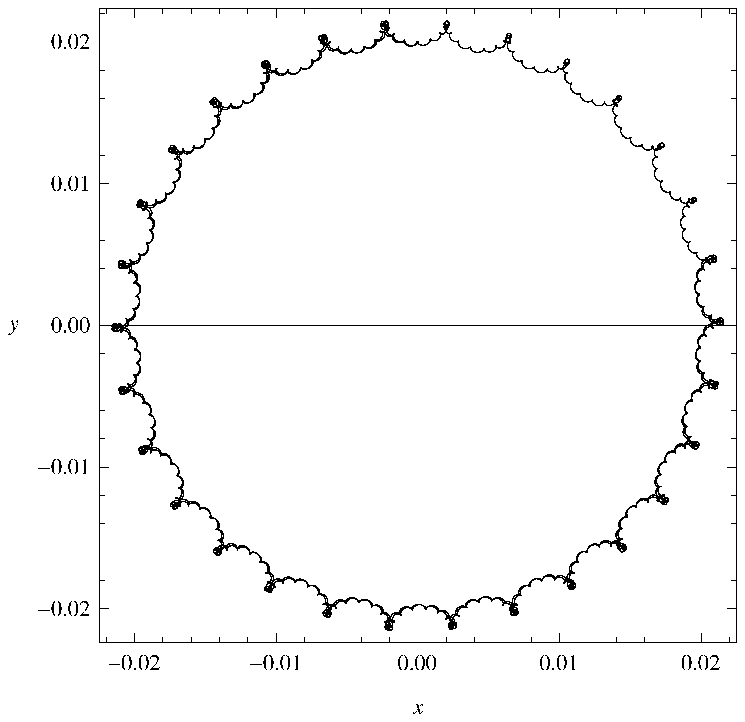}            
\par  
\end{centering} 
\vspace*{0.4cm}
\begin{centering} 
\includegraphics[width=6.5cm]{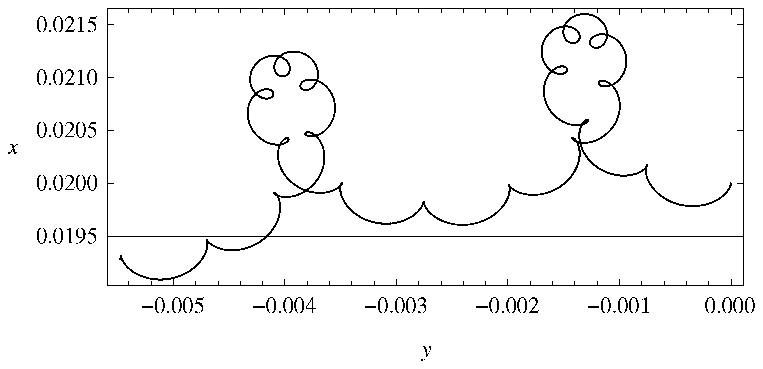}  \hspace*{1.0cm}  
\includegraphics[width=6.5cm]{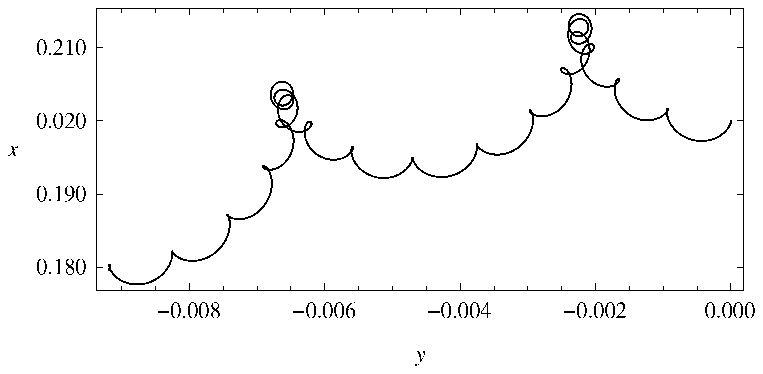}     
\par
\end{centering} 
\vspace*{0.2cm}
\caption{Top left: Numerical solution for the normal vector $\bm{\hat{\mathrm{n}}}$ 
   to the plane of the spinning ring, for 25 orbits about the black hole.
Note the nutation, upon which smaller precessional effects are superimposed.
The parameters used for this simulation are $M=1.5\times10^{7}\; M_{\odot}$
   (black hole), $m=1.5\; M_{\odot}$ (neutron star), $R=25r_{\mathrm{s}}$,
   where $r_{\mathrm{s}}=2GM/c^{2}=4.4\times10^{12}\,\mathrm{cm}$, and
   $\Omega=\sqrt{GM/R^{3}}=3.8\times10^{-5}\,\mathrm{rad\, s^{-1}}$,
   with the initial conditions $\alpha\left(t=0\right)=0.02\,\mathrm{rad}$,
   $\beta\left(t=0\right)=\pi/2$, $\gamma\left(t=0\right)=0$, $\dot{\alpha}\left(t=0\right)=\dot{\beta}\left(t=0\right)=0$,
   and $\dot{\gamma}\left(t=0\right)=\sqrt{2Gm/a^{3}}=3.8\times10^{-4}\,\mathrm{rad\, s^{-1}}$,
   where $a=2\; R_{\odot}$. 
 Bottom left: Magnification of solution for first
   full orbit about the black hole, illustrating the precessional effects within the nutation. 
Top right and bottom right: 
The corresponding results from pure Newtonian consideration are shown for comparison. 
They are different from the results obtained from the relativistic derivations.  
The Square Kilometer Array (SKA) is expected to achieve timing precision at the level of $\sim 100~{\rm ns}$ at 1.4~GHz for 10~min integration  
  for millisecond pulsars with normal brightness \citep{Liu11}. 
The differences in the nutation and precession shown for the two cases can be distinguished,  
  if one of the neutron stars in the DNS system is a millisecond pulsar.    
 } 
\label{nutation} 
\end{figure}

The numerical solution for the spatial position of $\bm{\hat{\mathrm{n}}}$ for 25 orbits 
  about the black hole, as seen from the positive $z$-axis, is shown in Fig.~\ref{nutation}.
Nutational behaviour is clearly evident, with other precessions superimposed.
For this case, the orbital motion of the ring about the black hole is approximately 3 percent relativistic; 
  that is, $d\tau\approx0.97dt$.
However, the importance of relativistic orbit-orbit coupling 
  in dictating the dynamics of the system is much higher; 
  assuming $\dot{\gamma}\gg\dot{\beta}$, we find for this system that 
\begin{equation}
    \left|\frac{V_{SL}}{V_{\mathrm{tidal}}}\right|\sim 6   \frac{\dot{\gamma}  \Omega  R^{2}}{c^{2}}\approx1.2   \   .
\end{equation}    
Note that precessions and nutations of the internal orbits of the neutron stars in the DNS system revolving around a black hole occur 
  even in a pure Newtonian consideration. 
However, the dynamics in the relativistic and the pure cases are qualitatively different, 
 in terms of the precession frequencies and in terms of the interplay between nutation and precession, 
  which can be seen in Fig.~\ref{nutation}.


\begin{figure} 
\noindent 
\vspace*{0.4cm}  \\
\begin{centering} 
\includegraphics[bb=100bp 0bp 688bp 154bp,clip,width=16.5cm]{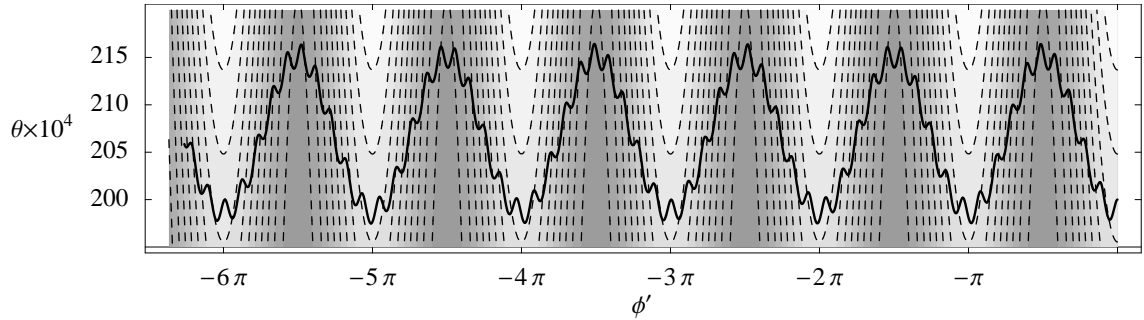}
\par
\end{centering} 
\caption{ 
Phase trajectory of $\bm{\hat{\mathrm{n}}}$ (solid line), 
  the normal vector to the spinning ring, plotted in co-rotating coordinates $\theta$ and $\phi'=\phi-\Omega t$ 
  for three orbits about the black hole. 
Contours (grayscale, dashed lines) indicate the effective potential energy $V_{\mathrm{tidal}}+V_{\mathrm{cen}}$,
  with lighter tones indicating higher potential energy.
} 
\label{comoving} 
\end{figure}


Greater intuition for the effect of the tidal force in governing the behaviour of the DNS system (the spinning mass ring) 
   may be obtained by plotting $\bm{\hat{\mathrm{n}}}$
   within the effective potential energy $V_{\mathrm{tidal}}+V_{\mathrm{cen}}$
   in co-rotating coordinates $\theta,\,\phi'$. This is done in Fig.~\ref{comoving}
   for three orbits about the black hole. 
Note that the ring's oscillations are in phase with the frequency of the rotating effective potential,
   a resonance suggested by our calculations in Eqs.~\ref{omega0theta} and~\ref{omega0phi}. 
The ring is driven to low values of $\theta$ at integer multiples of $\pi$ by the potential peaks and swings 
  to higher values of $\theta$ when it is energetically permissible to do so, i.e. at half-integer values of $\pi$.  

So far, we have not included the gravitational radiation loss in our analyses. 
For this system, the effects of gravitational waves on the dynamics are actually negligible. 
This can be understood as follows.   
The emission of gravitational radiation would impact the system on timescales of   
  $t_{\mathrm{grav}}\sim c^{5}a^{4}/G^{3}m^{3}$  or longer \citep{Misner77,Fang83}. 
However, the phenomena of the system operate at frequencies 
  on the order of $\Omega=\sqrt{GM/R^{3}}$ or $\Omega_{SL}\sim GM\Omega/Rc^{2}$.
For the astrophysically-motivated parameters used above, 
  $\Omega t_{\mathrm{grav}}\sim10^{13}$ and $\Omega_{SL}t_{\mathrm{grav}}\sim10^{12}$. 
The radiative loss timescales of gravitational radiation are therefore much larger than any dynamical timescales that we study.

\subsection{Remarks}  

Strongly bound gravitational systems often exhibit intriguing and complex dynamical phenomena. 
In particular, gravitational spin-orbit coupling and the precessional effects it engenders 
  provide an intriguing arena upon which post-Newtonian effects such as gravito-magnetism come into play.  
Experimentally confirmed to $\sim0.1$\% precision by lunar-ranging data \citep{Murphy07}, 
  detected in binary-pulsar systems \citep{Stairs04, Breton08},   
  and recently investigated by the Gravity Probe B experiment \citep{Everitt11}, 
  the spin-orbit interaction suggests deep analogies between gravitation
  and many other physical processes 
  \citep[see e.g.][for a discussion of experimental tests of the gravitational spin-orbit effect]{Blau11}. 
Across many physical systems, 
  the phenomenon of energetic coupling between different components of the angular momentum 
  is of paramount importance. 
More specifically, the coupling between orbital and intrinsic angular momenta is an important effect 
  in both quantum and classical systems. 
In quantum physics, the spin-orbit correction to the Hamiltonian is crucial 
  to understanding the fine structure of atomic spectra, 
  while in optics and condensed matter physics, respectively,
  examples include the case of polarization of light \citep{Niv08, Bliokh09}
  and the Josephson current \citep{Dell07} In classical physics,
  spin-orbit coupling occurs in fields from Maxwellian electrodynamics \citep{Jackson99} to astronomy, 
  in the case of tidal locking \citep{Escribano08}. 
The general relativistic problem of the spinning point particle is treated perturbatively 
  for orbits about a Kerr black hole in \cite{Singh08}.
A problem related to precession, 
  involving net rotational and translational displacements produced by cyclic motion of quasi-rigid bodies 
  (analogous to swimming) on curved manifolds, is discussed in \cite{Wisdom03}. 

For extended nonspherical bodies orbiting a central mass, tidal forces create additional precessional, nutational, and tumbling effects
   \citep[see e.g.][]{Mashhoon06}. 
In the non-inertial reference frame of the orbiting body, 
  fictitious forces must be taken into account to describe the evolution of the orientation. 
The behaviour of an orbiting body is governed by the combination of spin-orbit coupling, tidal forces, 
  and non-inertial effects, and cannot be described by deSitter precession alone. 
This wide and complex range of gravitational phenomena would occur in astrophysical systems   
  such as a DNS system infalling into a massive black hole. 
We have shown that the orbital revolution of the neutron stars in the DNS system  
  is dynamically a spinning mass ring of matter, of mass $2m$ and diameter $a$, 
  in orbit at radius $R$ about a central black hole with mass $M$. 
When the neutron stars are revolving around each other, 
  the DNS system constitutes a (nonspherical) gyroscope.  
The fixed radius $a/2$ in the ring representation of the DNS system 
  forms a holonomic constraint in the Lagrangian analysis.  
As a consequence, internal restoring forces counteracting the tendency of the ring to flex or stretch induced by the tidal interaction 
  do not impact the dynamical behaviour of the ring. 
Thus, the behaviour of a rigid mass ring orbiting around a gravitating object models the relevant effects in relativistic astrophysical systems,  
  such as a DNS system that itself is in orbit about a supermassive black hole in a hierarchical binary-structured three-body configuration. 

In general, a non-rotating black hole would not give rise to an explicit magnetic term under the GEM prescription 
 \citep[for the discussion of black hole gravito-magnetism see e.g.][]{Punsly01}. 
The hierarchical binary-structured three-body configuration 
  consisting of a DNS system orbiting around a non-rotating black hole is a unique astrophysical setting 
  in which GEM effects arise without invoking black hole rotation (Kerr spacetime).  
The revolution of the DNS system around the central object and the revolution of the neutron stars around each other 
  give rise to mass currents, producing magnetic moments and their magnetic-type coupling.  
This situation is analogous to induced magnetism. 
In the reference frame of the rigid ring, the black hole appears as a mass current 
  (similarly to the manner in which the proton appears as a charge current to the electron in the H atom in the electromagnetic problem). 
Hence, in the DNS system's reference frame, a gravito-magnetic field is present. 
When calculating the potential in this reference frame and changing to the centre of mass frame, 
  Thomas precession will appear as a consequence.  

Now, the remaining question is: what are the astrophysical consequences?  
Pulsars are high-precision spinning gyroscopes. 
They have been considered as useful tools to probe the spacetime properties 
  of the massive central black hole of our Galaxy \citep[see][]{Wex99,Liu12}.  
Moreover, they can also be used to investigate various general relativistic spin-orbit couplings \citep{Wex99,Singh05,Singh08,Iorio12}.  
Like pulsars, DNS systems are also high-precision spinning gyroscopes.  
Hence, DNS systems can also be used in experiments on gravity around black holes. 
In addition, as DNS systems are not point masses, 
  they can probe additional phenomena driven by the tidal gravitational field  
  and coupling caused by the mass current loop within the DNS systems.   
Although the chance that a DNS system with a pulsar is currently in close orbit with the massive black hole in the Galactic Centre 
 may not be substantial,   
 as future radio telescopes, in particular the Square Kilometer Array (SKA),  
 will be able to search for pulsars around central black holes in galaxies beyond the Local Group \cite[see e.g.][]{Smits09,Ridley10}, 
 such hierarchical three-body systems could be found in some external galaxies. 
 
We note that the analyses in this work are not restricted to DNS systems around massive black holes.   
They are applicable also to any hierarchical three-body systems 
  which contain a massive black hole and a tightly bound double compact object system    
  with various combinations of black holes and neutron stars, 
  e.g.\ stellar-mass black hole pairs (smBH-smBH), stellar-mass black hole and neutron star binaries (smBH-NS), 
         and intermediate-mass black hole pairs (imBH-imBH).  
The central 0.1~pc of galaxies similar to the Galaxy 
  may contain a few thousand stellar mass black holes and several hundred neutron stars 
    \citep[Table 1,][]{Hopman06}. 
Taking that roughly 10\% of neutron stars would be in binary systems \citep{Antonini12}, 
  one may expect about 10 DNS within 0.1~pc from the centres of these galaxies.   
 This is an optimistic estimate, as binaries in galactic centres are subject to various 2-body, 3-body and other non-linear scattering processes,  
   which would lead to binary disassociation and evaporation \citep[see e.g.][]{Hopman09, Perets09, Antonini12}.            
Nevertheless, DNS and other compact binaries will exhibit complex orbital dynamical properties due to orbit-orbit coupling 
  and some of these systems would exist in the central regions of galaxies.   
Tightly bound double compact object systems are gravitational wave sources and their complex orbital precession and nutation 
  will certainly have observational consequences in the properties of the gravitational waves from these systems. 
The discovery of pulsars in smBH-NS systems around a massive black hole \citep[see][]{Faucher11} would provide an opportunity for 
  an elegant double-test of general relativity, i.e.\ comparing gravitational wave signatures with pulsar-timing from the system.

\section{Conclusions} 

The behaviour of a DNS system orbiting around a massive black hole is governed in part 
  by coupling between the orbits of the neutron stars in the DNS system and the orbit of the DNS system around the black hole.  
The orbital motion of the neutron stars in the DNS systems drives a looped mass current, 
  inducing gravito-magnetism.   
Because of holonomic constraints,  
  the gravitational pull between the neutron stars does not contribute explicitly to the Lagrangian governing the system dynamics, 
  so that a tightly bound DNS system is effectively a rigid spinning mass ring.  
Working in this approximation framework, we have constructed the Lagrangian 
  and extracted the equations of motion for the system. 

The dynamical behaviour of this hierarchical three-body system illustrates various general relativistic effects. 
The tidal potential energy was derived via a parameterization of the tidal field as an effective potential. 
The general relativistic spin(orbit)-orbit coupling was derived for an arbitrary mass ratio, 
  by application of the GEM equations. 
The Lagrangian was formulated in the orbiting reference frame. 
It was shown that Coriolis and centrifugal effects collapse into the kinetic energy measured in the inertial frame, 
  while the tidal potential undergoes time-dependent coordinate rotation. 
The Euler angles are found to be related though a system of three coupled, nonlinear, time-dependent, transcendental differential equations.
Finally, the oscillatory behaviour of two restricted cases was examined in detail and a numerical solution of the general case was explored.

The motion of spinning bodies of finite extent, such as the DNS systems, within a gravitational field contains many sources of rich complexity. 
The system, though seemingly simple, holds many sources of interesting phenomena, among them tidal precession and oscillation, 
 spin-dependent nutation, and deSitter precession arising from the spin-orbit interaction. 
It is interesting to note the similarity of the results for both the gravitational quadrupole potential and gravitational spin-orbit coupling 
  between the corresponding quantum and classical systems.

\section*{Acknowledgments} 
We thank the referee for constructive comments and 
  for helpful suggestions in obtaining more reliable estimates for the compact objects and DNS in the centres of galaxies.

\bibliographystyle{mn2e}
\bibliography{Biblio}

\bsp 

\label{lastpage}
\end{document}